\documentclass[preprint,12pt]{elsarticle}

\pdfoutput=1
\usepackage{hyperref,amsmath,amsfonts,mathrsfs}
\usepackage{epstopdf}
\usepackage{tikz}
\usepackage{float}
\usepackage{flafter}
\usepackage{graphics}
\usepackage{graphicx}
\usetikzlibrary{shapes.geometric, arrows}

\journal{Theoretical Computer Sciences}

\begin{document}

\newtheorem{fact}{Fact}
\newtheorem{defn}{Definition}
\newtheorem{ft}{FootNote:}
\newtheorem{conj}{Conjecture}
\newtheorem{theo}{Theorem}
\newtheorem{lemma}{Lemma}
\newtheorem{prop}{Proposition}
\newtheorem{corr}{Corollary}
\newtheorem{abst}{Abstract}
\newtheorem{claim}{Claim}
\newtheorem{obs}{Observation}
\newtheorem{example}{Appendix}
\begin{frontmatter}



\title{There is a Hyper-Greedoid lurking behind every   Accessible  Graphical  Search Problems Solvable in Polynomial Time: $P \not = NP$}


\author{Kayibi Koko-Kalambay}

\address{School of Mathematics. Department of Sciences\\University of Bristol}

\begin{abstract}
Let $X$ be a set and let  $\mathcal{I}$ be a family of subsets of $X$.  A \textit{greedoid} is a pair $(X,\mathcal{I})$, where $\mathcal{I}$ satisfies the following axioms.

\begin{enumerate}
\item[G1.] \textit{Accessibility Axiom:} If $I \in \mathcal{I}$, then there is an element $x \in I$ such that $I - x\in \mathcal{I}$.
\item[M2.] \textit{Exchange Axiom:} If $I_{_{1}} \,\,\textrm{and}\,\, I_{_{2}}$ are elements of $\mathcal{I}$, and $|I_{_{1}}| > |I_{_{2}}|$,  then there is an element $ x \in I_{_{1}}  - I_{_{2}}$ such that $I_{_{2}} \cup x \in \mathcal{I}$.
\end{enumerate}

Let  $G(E,V)$ be an isthmus-less  labelled connected   graph  with  edge-set $E$ and vertex-set $V$. Let $G[X]$ denote the graph whose edge-set or  vertex-set is $X$.
Given an input  $G[X]$, 
  a \textit{graphical  search problem} associated with a predicate  $\gamma$   consists of  finding a subset $Y$  such that     $Y \subseteq X$, and $Y$ satisfies the condition  $\gamma$ in $G[X]$.   The set $Y$ is a \textit{solution} of the problem.  We denote such a problem as $\Pi(G[X], \gamma)$,  and we let  $\hat \Pi(G[X], \gamma)$ to denote the \textit{decision problem} associated with $\Pi(G[X],\gamma)$. \\
  
  A \textit{sub-solution} of $\Pi(G[X],\gamma)$ is a subset $Y'$ such that  $Y' \subset X$,  $Y'$ is  not a solution  of the problem  $\Pi(G[X],\gamma)$, but  $Y'$ is   a solution  of the problem  $\Pi(G[X'],\gamma)$, where  $X' \subset X$, and $G[X']$  is a \textit{contraction-minor} of $G[X]$.\\
  
   To each graphical search problem $\Pi(G[X],\gamma)$, we associate   the set system $(X, \mathcal{I})$,  where  $\mathcal{I}$ denotes the set of all the solutions and  sub-solutions of $\Pi(G[X],\gamma)$. \\

Given  a graphical search problem $\Pi(G[X],\gamma)$, we  relax  the Exchange Axiom M2 of greedoids  to define \textit{hyper-greedoids}     as follows. 
\begin{enumerate}  
\item[M2'.]  \textit{Augmentability:} If $I \in \mathcal{I}$, and $I$ is a sub-solution,  then there is a polynomial time computable function $\kappa$ from $\mathcal{I}$ to  $\mathcal{I}$ and   an element $x \in X - \kappa(I)$ such that $\kappa(I) \cup x \in \mathcal{I}$. (Note that the function  $\kappa$ is just an extension  of Edmond's Augmenting Path Algorithm. \cite{edmonds1} )
\item[G1.] \textit{Accessibility:}   If $I \in \mathcal{I}$, then  there is an element  $x$ in $I$ such that  $I-x \in  \mathcal{I}$. 
\end{enumerate}

 Let  $G$ be an isthmus-less labelled connected graph and let MISP be the problem that  consists of finding a \textit{maximal independent  set of vertices}   of  $G$.    We show in Lemmas \ref{MISAugmentable} and \ref{MISAccessible} that  the family of all the solutions and  sub-solutions of MISP satisfies Axioms G1 and M2', with $\kappa$ being the identity function.  By using the fact that the decision problem associated to MISP is in the computational class P-Complete,  we  show in Theorem \ref{MainTheorem} that  a graphical  search problem $\Pi(G[X],\gamma)$ that satisfies G1  is soluble in \textit{polynomial time} if and only if   the set system $(X, \mathcal{I})$  satisfies   M2', the Augmentability Axiom.  \\

The  $\hat{HCP}$ decision  problem  consists of finding out whether or not  a graph $G$ contains a  \textit{Hamiltonian cycle}.  The  problem  $\hat{HCP}$ is  in the computational class   $NP$-complete, and  the associated  search problem $HCP$  is  accessible (satisfies G1) but does not satisfy  Axiom M2', as we show in Lemmas \ref{HamiltonianAccessibility}, \ref{HCPnotAugmentable} and \ref{HCPnotSlowAccessible}.  We  thus conclude in Corollary \ref{pnotnp}
 that the computational complexity  class $P$ is different from the computational  complexity class $NP$. That is,   $P \not = NP$.\\

\end{abstract}

\begin{keyword}Algorithms \sep Greedoid  \sep Matroid \sep Generalised Greedy Algorithm \sep Polynomial time \sep P versus NP .

\MSC  05C99 \sep 11K99  

\end{keyword}

\end{frontmatter}



\section{Introduction}

\subsection{Computational Complexity Requisites}\label{Computational Complexity Requisites}
This section concerns the key notions of Computational Complexity Theory needed for the  proof  of Theorem \ref{MainTheorem}.  An \textit{algorithm} $\mathcal{A}$  is a sequence of instructions that transforms an input $X$ into an output $Y$. The \textit{run time} of $\mathcal{A}$ with input $X$, denoted  $t_{_{\mathcal{A}}} (X)$,  is  the number of steps in the computation of $\mathcal{A}$ on input $X$. The time  $t_{_{\mathcal{A}}} (X)= \infty$ if this computation never halts. Let $T_{_{\mathcal{A}}} (n)$ be  the worst case run time of $\mathcal{A}$ on all inputs of size $n$. That is, 
\[
T_{_{\mathcal{A}}}(n)=\max\{t_{_{\mathcal{A}}}(X):|X|=n\}. 
\]

An algorithm   $\mathcal{A}$  runs in \textit{polynomial time} if there exists  a real number $k$ such that for all $n$, 
\[
T_{_{\mathcal{A}}} (n) \leq n^{k}. 
\]

That is, the number of steps taken by the computation is bounded above by a polynomial on the size of the input. 
An algorithm   $\mathcal{A}$  runs in \textit{exponential time} if there exists a  positive real  number $k > 1$ such that for all $n$, 
\[
T_{_{\mathcal{A}}} (n) \geq k^{n}. 
\]

A \textit{decision problem} is a problem that takes some input  $X$, and outputs "yes" or "no". Roughly speaking,  the class $\mathcal{P}$ consists of all those decision problems  that can be solved by an algorithm that runs in an amount of time that is polynomial in the size of the input. 

The \textit{class $\mathcal{NP}$} consists  of all those decision problems whose positive solutions can be verified in polynomial time given the right information, called  a   certificate $Y$. To each  $\mathcal{NP}$ decision problem is associated   a \textit{search problem}, which is, given a string $X$, find a string $Y$ such that $Y$ is a certificate of membership of $X$ in some class $L$ (or determine that no such certificate exists).\\

{\bf Definitions 1.1.} \label{reducibility} A decision problem $\hat \Pi$ is \textit{reducible}  to another problem $\hat \Pi'$ 
if,  given an instance   $X$ of $\hat \Pi$,   there is an algorithm  $\phi$ that transforms $X$  into an instance $X'$ of  $\hat \Pi'$ and an algorithm  $\psi$  that  transforms each solution of $\hat \Pi'$  into a solution of $\hat \Pi$, such that $Y$ is a solution of $X'$ in  $\hat \Pi'$ if and only if $\psi(Y)$ is a solution  of $X$ in $\hat \Pi$. This means that a solution to the $\hat \Pi'$ problem provides a solution for the problem $\hat \Pi$.  

A decision problem $\hat \Pi$ is \textit{complete} for the class $\mathscr{C}$ if it is in $\mathscr{C}$, and all the problems in $\mathscr{C}$ can be reduced to it in an appropriate manner. Or, given an algorithm $\mathcal{A}$ for a  problem complete for $\mathscr{C}$, any problem in $\mathscr{C}$  can be solved by an algorithm $\mathcal{B}$ that uses $\mathcal{A}$ as sub-routine. A decision problem $\hat \Pi$ is \textit{hard} for the class $\mathscr{C}$ if  all problems in $\mathscr{C}$ can be reduced to it  in an appropriate manner. \\


The notions of \textit{$\mathcal{NP}$-complete}  and \textit{$\mathcal{P}$-complete} problems  are essential for what follows in the present paper. $\mathcal{NP}$-complete problems are the set of problems to each of which any other $\mathcal{NP}$ problem can be reduced in polynomial time, and whose solution may still be verified in polynomial time. Similarly, $\mathcal{P}$-complete problems are the set of problems to each of which any other $\mathcal{P}$ problem can be reduced in polynomial time.\\
 
The \textit{Boolean Satisfiability Problem}, denoted  $\hat{SAT}$, is the problem of determining whether there exists an interpretation that satisfies a given Boolean formula. That is, given a Boolean formula,  can the variables  be consistently assigned the values `true' or `false' in such a way that the formula evaluates to `true'. The following folklore  results of Computer Sciences will be used throughout the present paper.
 
 \begin{theo} \cite{cook2}
  $\hat{SAT}$ is $\mathcal{NP}$-complete.
 \end{theo}
 
\begin{theo}\label{SATeqHatSAT}\cite{bellare}
The decision problem $\hat{SAT}$ is in $\mathcal{P}$ if and only if the search problem SAT is solvable in polynomial time.
\end{theo}

Consider the graph  $G(V,E)$, where $V$ is the vertex-set and $E$ is the edge-set. A \textit{Hamiltonian cycle} of $G$ is a cycle that contains all the vertices of $G$, while a \textit{Hamiltonian path} is a path that passes through all the vertices of $G$ exactly once.  The \textit{Hamiltonian Cycle Problem}, denoted $HCP$,  consists of finding such a Hamiltonian cycle. We denote by $\hat{HCP}$ the decision problem associated with $HCP$.

\begin{theo} \cite{cook2}\label{hamiltonCycleNP}.
$\hat{HCP}$ is $\mathcal{NP}$-complete.
\end{theo}

\begin{theo}\label{HCeqHatHC}
The decision problem $\hat{HCP}$ is in $\mathcal{P}$ if and only if the search problem HCP is solvable in polynomial time.
\end{theo}


An \textit{Acyclic Boolean  Circuit}   is a collection of gates (and, or, not) and wires that performs a mapping from Boolean inputs (0,1) to Boolean outputs (0,1), and contains no loops (always feeds forward). 
Given an Acyclic Boolean Circuit with several inputs
and one output and a truth assignment to the inputs, the \textit{Circuit Value Problem} (CV) consists of finding the value of
the output.

\begin{theo} \cite{ladner1}\label{circuitValueProblemEasy}
$\hat{ CV}$ is $\mathcal{P}$-complete.
\end{theo}

Consider the graph  $G(V,E)$, where $V$ is the vertex-set and $E$ is the edge-set.  An \textit{independent set of vertices} is a subset of vertices $U \subseteq V$ such that no two vertices in $U$ are adjacent. An independent set  $U$ is \textit{maximal} if no vertex can be added to it without violating independence. An independent set is \textit{maximum} if it has the largest cardinality  amongst all the independent sets.  The \textit{Maximal Independent Set} problem, denoted $MISP$, is the problem that consists of finding a maximal independent set of the graph $G$, while the \textit{Maximum Independent Set } problem, denoted $MaxISP$, consists of finding an independent set of the greatest cardinality. We denote by $\hat{MISP}$ the decision problem associated with  $MISP$.  


\begin{theo} \cite{cook0}\label{maximalISEasy}
$\hat{MISP}$ is $\mathcal{P}$-complete.
\end{theo}

By Theorems \ref{hamiltonCycleNP} and \ref{maximalISEasy},  there are  decision problems on graphs that are $\mathcal{P}$-complete or $\mathcal{NP}$-complete. That is, there are `prototypical' problems in $\mathcal{P}$  and $\mathcal{NP}$ that can be expressed in terms of graphs. In other words, every decision problem in $\mathcal{P}$ is the Maximal Independent Set  Problem ($\hat{MISP}$) in disguise, while every decision problem in $\mathcal{NP}$ is the Hamiltonian Cycle Problem  ($\hat{HCP}$) in disguise thanks to the conversion via a function $\phi$. \\

\subsection{ P   versus NP Problem }

The `P versus NP problem'  consists of showing    whether every algorithmic problem with efficiently verifiable solutions have efficiently  computable solutions.\\

 The problem was posed by Edmond  in 1967, and had prompted  far reaching researches in Theoretical Computer Sciences.  One of the avenues of research to solve the  `P versus NP'  Problem is the P-isomorphism Conjecture, by Berman and Hartmanis, which states that  any two NP-Complete sets   $L_{_{1}}$ and $L_{_{2}}$ are p-isomorphic to each other.  That is, there is a polynomial time computable, polynomial time  reduction F from $L_{_{1}}$ to $L_{_{2}}$ which is 1-1 and onto. That is, $F$ is a bijection between $L_{_{1}}$ and $L_{_{2}}$, and  thus, $L_{_{1}}$  and  $L_{_{2}}$  are essentially `copies' of one another. \\

The most interesting aspect of the conjecture is that, if it is true, then $NP\not= P$. Indeed, if NP = P, then, even finite sets would be NP-Complete. But, a finite set cannot be isomorphic to a infinite set like SAT.\\

 We take a different approach based on an unpublished `conjecture'  by Dominic Welsh, which states  that  `there is a matroid lurking  behind ever every good algorithm'. So, we extend the axioms of greedoids to englobe some set systems consisting  of solutions and sub-solutions of \textit{accessible  graphical search problems solvable in polynomial time}.  \\

The fundamental intuition is based on the fact that  there are `prototypical'  $\mathcal{P}$-complete  and $\mathcal{NP}$-complete problems  that can be expressed in terms of graphs. Indeed, every decision problem in $\mathcal{P}$ is the Maximal Independent Set  Problem ($\hat{MISP}$) in disguise, while every decision problem in $\mathcal{NP}$ is the Hamiltonian Cycle Problem  ($\hat{HCP}$) in disguise. Thus, we only have to concern ourselves with finding the inherent combinatorial properties that make the $\hat{MISP}$ problem to be  $\mathcal{P}$-complete, and a solution of $\hat{HCP}$  to be easy to check but hard to find.  Hence, without loss of generality,  we may restrict ourselves on cases where $X$ is either the set of edges or vertices of an isthmus-less labelled connected graph to characterise completely the computational class $\mathcal{P}$.  \\

We then show that the $MIS$  Search problem satisfies the \textit{Augmentability Axiom}.  Augmentability  entails that, in the quest for a solution $Y$ of a Graphical Search Problem, if one starts from the empty set and moves from  one sub-solution to another sub-solution by augmentation (adding one element at a time), then, every move is a right move towards a solution $Y$, provided  every solution is accessible. Thus, there would be no backtracking, and a solution can be  found in polynomial time if each augmentation can be made in polynomial time.\\

 Conversely,  suppose that there is a sub-solution $Y'$  that is not augmentable, as we show for the search problem  $HCP$. Then, an algorithm searching for a solution by building it from the empty set  has to avoid  getting stuck at $Y'$. Thus, for every element  $x$ added iteratively, the algorithm has to check exhaustively  all the supersets of the sub-solution reached so far to anticipate which one is augmentable. Hence the algorithm would be exponential, and  in the worst cases, it  has to backtrack. Much of the present paper is about turning this intuition into a sound mathematical proof.\\

\subsection{ Matroids, Greedoids  and the Greedy Algorithm}

A graph  $G(V,E)$ is \textit{labelled} if its edges and vertices are indexed so that $V= \{v_{_{1}}, \cdots, v_{_{n}}\}$ and   $E= \{e_{_{1}}, \cdots, e_{_{m}}\}$.  A graph  $G$ is \textit{connected} if, given any two  of its vertices $v_{_{i}}$ and  $v_{_{j}}$, there is a path connecting them. A graph  $G$ is \textit{2-connected} if given any two of its edges $e$ and $f$, there is a cycle $C$ such that  $e$ and $f$ belong to $C$. We say  that $G$ is \textit{isthmus-less} if  every edge of $G$ belongs to some cycle.   Throughout the present paper, we assume $G$ to be an isthmus-less  labelled  connected graph. For a graph $G(E,V)$, the notation $V(G)$ and $E(G)$  stand for the sets of vertices of $G$ and edges of $G$, respectively.  A graph $H(E',V')$  is a \textit{minor}  of $G(E,V)$ if  $H(E',V')$  is  obtained from $G$ through a sequence of  \textit{deletions} and  \textit{contractions} of edges of $G$, as explained in  Appendix \ref{exampleContraction}. Let $G$ be a graph,  and let $X$  and $A$ be  sets.  Throughout the present  paper,   the notation $G/A$ stands for the \textit{contraction} by the set of edges in $A$,  while  the notation $G/(A-e)$ stands for the \textit{re-insertion} of the edge $e$ into the graph $G/A$, i.e., the operation  which consists of reversing the contraction by the edge $e \in A$, as explained in Appendix \ref{exampleContraction}. The notation   $X-A$   stands for the removal of the  elements of the set $A$ from the set $X$.   If  $A $ is a set and $a$ is an element of another set  $B$, then the notation $A \cup a$ stands for $A\cup \{a\}$.   \\ 

We follow closely the notations of \cite{anderson1, cook1, dasgupta} for Theoretical Computational Complexity, the notations of \cite{oxley1, welsh1} for Matroid Theory and the notations of \cite{bjorner1} for Greedoid  Theory. \\

Let $X$ be a set and  let $\mathcal{I}$ be a family of subsets of $X$.  We refer to elements of $\mathcal{I}$ as \textit{feasible sets}.
A \textit{simplicial complex} is a pair $(X,\mathcal{I})$, where $\mathcal{I}$ satisfies the following axiom.
\begin{enumerate}
\item[M1.]\textit{Heredity Axiom:} If $I \in \mathcal{I}$, then for all  $x \in I$, $I -x \in \mathcal{I}$.
\end{enumerate}

A \textit{matroid}  is a pair $(X,\mathcal{I})$, where $\mathcal{I}$ satisfies M1, the Heredity Axiom, and Axiom M2:

\begin{enumerate}
\item[M2.] \textit{Exchange Axiom:} If $I_{_{1}} \,\,\textrm{and}\,\, I_{_{2}}$ are elements of $\mathcal{I}$, and $|I_{_{1}}| > |I_{_{2}}|$,  then there is an element $ x \in I_{_{1}} - I_{_{2}}$ such that $I_{_{2}} \cup x \in \mathcal{I}$.
\end{enumerate}

A \textit{greedoid} is a pair $(X,\mathcal{I})$, where $\mathcal{I}$ satisfies  the  following axioms.

\begin{enumerate}
\item[G1.] \textit{Accessibility Axiom:} If $I \in \mathcal{I}$, then there is an element $x \in I$ such that $I - x\in \mathcal{I}$.
\item[M2.] \textit{Exchange Axiom:} If $I_{_{1}} \,\,\textrm{and}\,\, I_{_{2}}$ are elements of $\mathcal{I}$, and $|I_{_{1}}| > |I_{_{2}}|$,  then there is an element $ x \in I_{_{1}}  - I_{_{2}}$ such that $I_{_{2}} \cup x \in \mathcal{I}$.
\end{enumerate}


The \textit{ Greedy Algorithm} is a generalisation of the Kruskal Algorithm for finding $B$, a spanning tree of minimal cost of a graph.  Indeed, let $G$ be a graph with edge-set $X$  and let $\mathcal{I}$ be the collection of \textit{independent sets of edges} in  $X$. Suppose that each edge $e$ is assigned a weight $w$ such that for each subset $A \subseteq X$, $w(A)= \sum_{_{e \in A}} w(e)$. The Greedy Algorithm proceeds as follows.\\

\begin{enumerate}
\item[Step 1] Set $Y_{_{0}}= \phi$ and $i=0$.
\item[Step 2] If $X - Y_{_{i}}$ contains an element $e$ such that  $Y_{_{i}} \cup e \in \mathcal{I}$, choose such an element $e$ of minimal weight, let $Y_{_{i+1}}= 
Y_{_{i}} \cup e$ and go to Step 3. Else, let $Y_{_{i}}=B$ and go to Step 4.
\item[Step 3]  Let $i=i+1$ and go to Step 2.
\item[Step 4] Stop
\end{enumerate}

 We note that the algorithm  runs in polynomial time if  recognising that  $Y_{_{i}} \cup e \in \mathcal{I}$ can be done in polynomial time.\\
 
  Prior to Hassler Whitney \cite{oxley1}  defining matroids axiomatically as  generalisations of linear independence,   Boruvka Otakar \cite{nesetril}  used implicitly the  axioms of  matroids to justify the optimality of the greedy algorithm he proposed for finding a Minimum Spanning Tree of a graph. And, indeed, it happens that matroids are the only structures where the greedy algorithm outputs an optimal solution for all  weights functions $w$.  
 
\begin{theo} \cite{oxley1}

Let $\mathcal{I}$  be a collection of subsets of a set $X$.  Then  $(X, \mathcal{I})$   is a matroid if and only if $\mathcal{I}$ satisfies the following conditions:

\begin{enumerate}
\item $\emptyset \in \mathcal{I}$
\item If $I \in \mathcal{I}$ and $I' \subseteq  I$, then $I' \in \mathcal{I}$

\item For all weight functions $w: X \rightarrow \mathbb{R}$, the greedy algorithm produces a  member  $I \in \mathcal{I}$ of maximum weight.
\end{enumerate}
 \end{theo}

Later Korte and  Lovasz \cite{bjorner1}  observed that, in some cases,  it only suffices that $\mathcal{I}$  satisfies the Exchange and  Accessibility axioms  for the Greedy Algorithm  to output an optimal solution.  Thus, we have that whenever the  combinatorial structure of a search  problem  is a greedoid, there is an algorithm that solves the problem   in polynomial time.  The present paper attempts to relax the axioms of greedoids and  to reverse the above  implication. That is, the existence of a polynomial time algorithm  implies the existence of a   combinatorial structure that is a `natural extension' of a greedoid. \\

 The enfolding of this article is organised in three sections. In Section Two,  we define the feasible sets   of a computational  problem $\Pi(G[X], \gamma)$,  we define the  closure of feasible sets, and we  give some examples in the Appendices  to help understanding.  We then  present the main result in  Theorem \ref{MainTheorem},  its proof and its  main consequence in Corollary \ref{pnotnp} in Section Three. The proof   is divided in many Lemmas for the sake of convenience.

 \section{Main Definitions and  Results}
 
 \subsection{Feasible sets of $\Pi(G[X], \gamma)$: solutions and sub-solutions }

\vspace{5mm}

  {\bf Definitions 2.1. }\label{searchProblem}   A \textit{ Graphical   Search Problem}  associated with the predicate $\gamma$, denoted $\Pi(G[X],\gamma)$,  consists of finding a  subset  $Y \subseteq X$, where $X$ is either the edge-set or the vertex-set of a graph $G$, and  $Y$  satisfies the condition $\gamma$ in $G$.  We say that  $G[X]$  is the \textit{instance} or the \textit{input} of the search problem, or the search problem  is instanced on $G[X]$,   or the search problem is restricted to   $G[X]$, and we say that $Y$ is a  \textit{solution} or a \textit{basis} of $\Pi(G[X],\gamma)$.  The \textit{decision problem} associated with $\Pi(G[X],\gamma)$, denoted  $\hat \Pi(G[X], \gamma)$,  consists of finding whether or not  there is a solution $Y$, where $Y \subseteq X$ and $Y$ satisfies $\gamma$.\\

  Prototypical examples  of  graphical  search  problems are the STP, the MISP, the MMP and the HCP problems. The   STP, illustrated in Appendix \ref{exampleSTP1}, is the problem that consists of finding a \textit{spanning tree} of  the graph $G[X]$. That is, finding a   set $Y$ of edges that connects all the vertices of $G[X]$, but does not contain  a cycle. Thus $X= E$, the set of edges of $G[X]$.  The problem MISP, illustrated in  Appendix \ref{exampleMIS}, consists  of finding a \textit{maximal independent set}  of  vertices of  $G[X]$. That is, finding a set $Y$ of vertices that are not adjacent to each other and no other vertex can be added without violating independence. Here, $X= V$, the set of vertices of $G[X]$. The MMP, the Maximum Matching Problem, illustrated in Appendix \ref{exampleMATCHING}, consists of finding a  \textit{maximum matching}. That is, finding   a set $Y$  of edges of the greatest cardinality    such  that   no two of these edges have an endpoint in common in the  graph $G[X]$. Thus $X= E$, the set of edges of $G[X]$.    The HCP Problem, illustrated in Appendix \ref{exampleHCP},  is the problem that consists of finding a  \textit{Hamiltonian cycle} of the graph $G[X]$. That is, finding a cycle that is incident to  all the vertices of $G[X]$. Thus $X= E$, the set of edges of $G[X]$.\\


\vspace{3mm}     

 
Given a graph $G(V,E)$,  a \textit{contraction-minor} of $G$, denoted $G/A$,  a \textit{deletion-minor} of $G$, denoted $G\setminus B$, and a  \textit{minor} of $G$, denoted $G\setminus B/A$,  where $A, B  \subseteq E$, are explained in Appendix \ref{exampleContraction}. A \textit{partial Hamiltonian cycle} of  $G(E,V)$  is a Hamiltonian cycle of a \textit{minor} of   $G(E,V)$. Lemma  \ref{HCPSolution}, whose proof is given in Appendix   \ref{proofOfEquivalence}, shows  that if $C'$ is a partial hamiltonian cycle of  $G(E,V)$, then there is a contraction-minor $H$ of   $G(E,V)$ such that $C'$ is a hamiltonian cycle in $H$.

 \begin{lemma}\label{HCPSolution}
Given a graph $G(E,V)$,  every  partial Hamiltonian cycle $G$  is a Hamiltonian cycle of a contraction-minor of $G$. 
\end{lemma}

Since the HCP Problem is a NP-Complete, that is, it encodes every other problem in the class NP, Lemma  \ref{HCPSolution} prompts the following generalisation, given as Definition 2.2.

\vspace{3mm}

{\bf Definitions 2.2. }\label{subsolution}   

  Given the  problem  $\Pi(G[X],\gamma)$, we say that $Y'$ is a \textit{sub-solution}  of $\Pi(G[X],\gamma)$   if 

  \begin{itemize}  
  \item  $Y'$ is not a solution of $\Pi(G[X],\gamma)$,  and 
   \item   $Y'$ is a solution of  $\Pi(G[X'], \gamma)$, where  $X' \subset X$, and   $X'$ is   either the edge-set or the vertex-set of  a contraction-minor $G/A$ of $G[X]$. 
   \end{itemize} 

  The graph $G[X']$ is said to be a \textit{sub-instance}.  A \textit{feasible set} of $\Pi(G[X], \gamma)$ is either a sub-solution or a solution (basis) of $\Pi(G[X],\gamma)$.   \vspace{3mm}
 
  Next come Lemma  \ref{MisIndependentSolution}  to Lemma  \ref{MaximumMatchingSolution}, the proofs of which are given in  Appendix  \ref{proofOfEquivalence}, which further justify our definition of feasible sets as the solutions of the problem $\Pi(G[X], \gamma)$ instanced on  contraction-minors of the graph $G[X]$.  Recall that a set $A$ of vertices of $G$ is \textit{independent} if the vertices in $A$ are  not adjacent to each other, and a set  $A$ of edges of $G$ is \textit{independent} if $A$ does not contain a cycle.

 \begin{lemma}\label{MisIndependentSolution}
There is a bijection between the set of independent sets of vertices of  $G(E,V)$ and the set of  feasible sets of the  Maximal Independent Set Problem $\Pi(V, \gamma)$. That is, every  independent sets of vertices $G$ is a feasible set of the  problem MISP, and  every feasible set of MISP on $G$  is an independent sets of vertices of $G$. 
\end{lemma}

\begin{lemma}\label{STPIndependentSolution}

There is a bijection between the set of  independent sets  of edges of  $G(E,V)$ and the sets of  feasible sets of the  Spanning Tree Problem $\Pi(E, \gamma)$. That is, every tree  of $G$ is a feasible set of the Problem STP and  every  feasible set is a tree of $G$.
\end{lemma}

\begin{lemma}\label{MaximumMatchingSolution}

There is a bijection between the \textit{matchings}  of   $G(E,V)$ and the sets of   feasible sets of the  Maximum Matching   Problem $\Pi(G[E], \gamma)$. That is, every matching  of $G$ is a feasible set of the Problem MMP and  every  feasible set is a matching of $G$.  
\end{lemma}

  We note that all the    decision  problems concerned in Lemmas   \ref{MisIndependentSolution},   \ref{STPIndependentSolution}, \ref{MaximumMatchingSolution}  and \ref{HCPSolution}   are  either P-complete, NP-complete or solvable in polynomial time. Thus, the  definition of sub-solutions in Definition 2.2 seems `intuitively'  natural as it fit  well with  some problems that `encode' all the polynomially computable and polynomially checkable  problems.   As a bonus from  Definition 2.2, we get that the   feasible sets of  the  Hamiltonian Cycle Problem (a prototypical NP-complete problem) and the feasible sets of the  Maximal Independent Set Problem (a prototypical P-complete problem)  are accessible (satisfies Axiom G1), as proved in Lemmas \ref{MISAccessible}  and \ref{HamiltonianAccessibility}.    \\

\vspace{3mm}

 A function $\kappa$ is said to be \textit{polynomial time computable} if its output can be reached in time that is polynomial in the size of the input.   A feasible set  $Y'$ is  \textit{augmentable}   if there is a  polynomial time computable function $\kappa$ from $\mathcal{I}$ to $\mathcal{I}$  and    an element $x \in X - \kappa(Y')$ such that $\kappa(Y') \cup x$ is a feasible set.  If the function $\kappa$ is the identity function, we say that $Y'$ is \textit{fast-augmentable}. That is,  there is  an element $x \in X - Y'$ such that $Y' \cup x$ is a feasible set.  A feasible set $Y'$ is \textit{accessible} if $Y'-y$  is a feasible set for some $y \in Y'$. We say  that the problem $\Pi(G[X], \gamma)$ is \textit{accessible}  if every feasible set is accessible, and the problem $\Pi(G[X], \gamma)$ is \textit{augmentable}  if every sub-solution  is augmentable.\\

  Suppose that $I'$ and $I''$ are two feasible sets. We write $I'' \unlhd I' $  if $I'' \subset  I'$ and $|I''| = |I'|-1$.  Suppose that every feasible  set $I$ of the problem $\Pi(G[X], \gamma)$  is \textit{accessible}. Then   there is  a chain   $\emptyset \unlhd I^{^{(1)}} \unlhd I^{^{(2)}} \unlhd  \cdots  \unlhd I$, where  every $I^{^{(i)}}$ is a feasible set. Such  a chain is called a \textit{chain of accessibility}, which entails that there is a steady  path from  $I$ to the empty  set.

\vspace{3mm}

 {\bf Definitions 2.3. }\label{closure} In what follows, we recall that the notation  $H[X]$ stands for the graph $H$, whose edge-set or vertex-set is $X$.\\
 
   For a feasible set $Y'$ of the graphical search problem  $\Pi(G[X], \gamma)$,  a \textit{closure} of $Y'$, denoted $cl(Y')$,  is $Y'\cup A$,   a maximal superset of  $Y'$ such that   
  
  \begin{itemize} 

    \item $Y'$ is a solution of $\Pi(H[ Y' \cup  A], \gamma)$, and
    \item   $H[Y' \cup A]$ is a contraction-minor of the graph $G[X]$. 
 \end{itemize}   
    
    That is, 

\[
cl(Y')=Y' \cup  A,
\]  
  where $A$ is a maximal set of   elements $x \in X-Y'$  such that $ Y'$ is a solution (basis) of the problem $\Pi$ restricted to  the graph $H[Y' \cup  A]$,  where  $H[Y' \cup  A]= G/B$ for some $B \subseteq E$. 
  \\
  
   Once again,   the example of the Spanning Tree Problem, given in Appendix \ref{exampleSTP1},  illustrates the fact that our definition of closure is just a natural  extension  of the  closure of a greedoid.   \\

   As a  contraction-minor,  the graph  $H[Y' \cup A]$  can be  constructed from $G$ as follows:
  
  \begin{enumerate}
 
  \item [Step 1:] Consider a contraction-minor $ H_{_{0}}= G/ \{ e_{_{1}}, e_{_{2}}, \cdots, e_{_{r}}\}$ such that $Y'$ is a solution of $\Pi(H_{_{0}}, \gamma)$. Such contraction-minor  exists since $Y'$ is a sub-solution. Let   $H[cl(Y')]= H_{_{0}} $.\\

  \item [Step 2:]   If there is an edge $e_{_{k}} \in \{ e_{_{1}}, e_{_{2}}, \cdots, e_{_{r}}\}$  such that re-inserting $e_{_{k}}$ yields a contraction-minor 
  $G/ \{ e_{_{1}}, e_{_{2}}, \cdots, e_{_{k-1}}, e_{_{k+1}},\cdots , e_{_{r}}\} $   and $Y'$ is still a solution of  the problem restricted to $ G/ \{ e_{_{1}}, e_{_{2}}, \cdots, e_{_{k-1}}, e_{_{k+1}},\cdots , e_{_{r}}\}    $,\\
   let  $ H[cl(Y')]=  G/ \{ e_{_{1}}, e_{_{2}}, \cdots, e_{_{k-1}}, e_{_{k+1}},\cdots , e_{_{r}}\}$, repeat  Step 2.  Else Stop. 
 \end{enumerate}

  



Notice that   $cl(Y')$ may not  be unique.  There may be many closures of $Y'$ of the form $Y' \cup P_{_{i}}$.  However, we have   that $cl(Y')$ is unique in MISP. \\

\begin{lemma}\label{MISClosureUnique} 

Let $Y'$ be a feasible set of the problem  MISP on the graph $G$. Then $cl(Y')$ is unique
\end{lemma}

{\bf Proof.}
Let $Y'$ be a feasible set of  MISP. That is, $Y'$ is a subset of  the set of vertices  that are not adjacent to each other in $G$.  Now,   $cl(Y')= Y' \cup  A$, where $A$  is the set of all  the vertices in $X - Y'$ that are connected to some vertex of $Y'$ in $G$. Hence, $cl(Y')$ is unique (as a set of vertices)  for all feasible sets $Y'$.                                         
\qed\\

We give an algorithmic construction of $cl(Y')$ for  MISP in Appendix  \ref{exampleMIS}, where we give some  important examples of closures of different graphical search problems.   The  example of MISP in Appendix \ref{exampleMIS}, and the example of HCP,  in Appendix \ref{exampleHCP}, 
are given since they are the two main problems this paper is concerned with. \\\\

  Now we give an important  series of lemmas   that will be much used in the proof of Theorem \ref{MainTheorem} and Corollary \ref{pnotnp}. All these lemmas concern  the Maximal Independent Set Problem (MISP), which is a  prototype of  $\mathcal{P}$-complete problem, and the Hamiltonian Cycle Problem (HCP), which is a  prototype of  $\mathcal{NP}$-complete problem.  
\vspace{3mm}

\begin{lemma}\label{MISAugmentable} 
Every sub-solution (feasible sets that are not bases) of the Maximal Independent Set Problem is augmentable. That is, MISP problem satisfies M2'. 
\end{lemma}

\begin{figure}[h]
\includegraphics[scale=0.42]{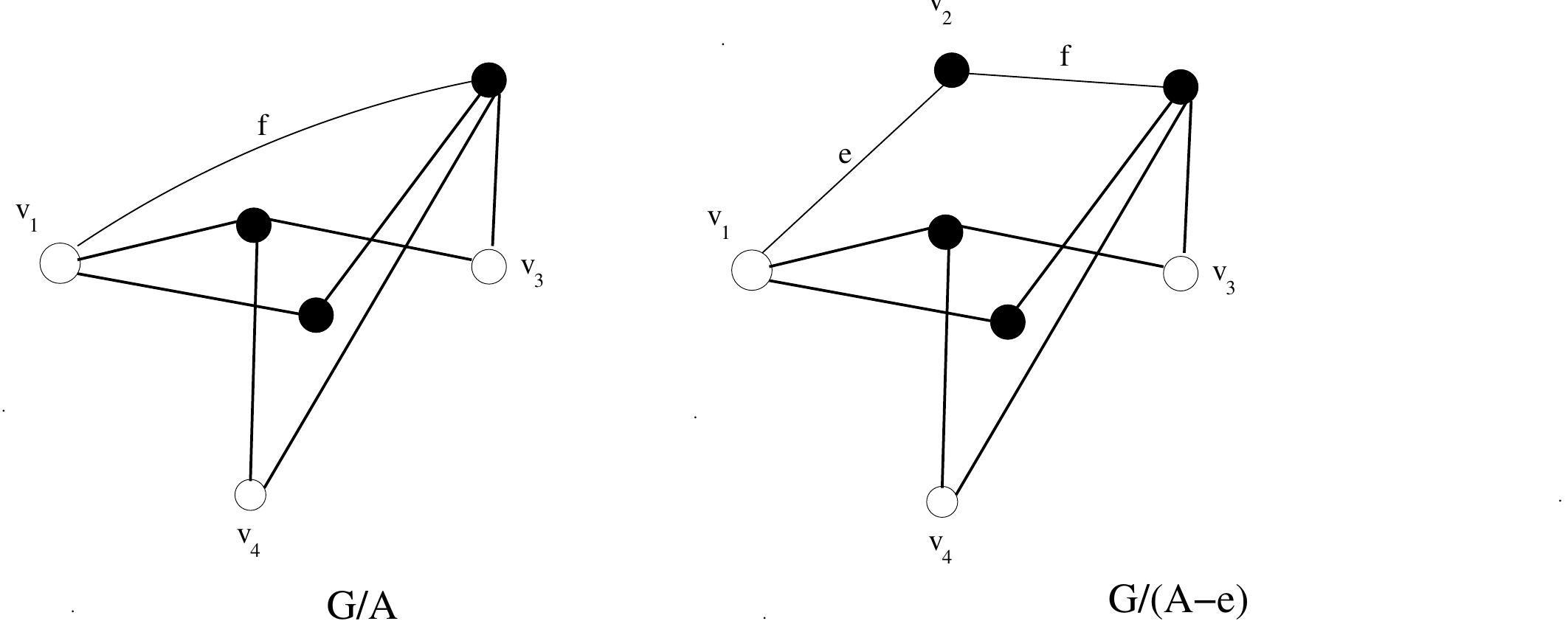}
\caption{ Suppose that $A= \{e\}$ so that  $G= G/(A-e)$. The set $\{ v_{_{1}},  v_{_{3}}, v_{_{4}}\} $ is a maximal independent set of the graph $G/A$. The vertex $v_{_{2}}$ can not  be added to it since $v_{_{2}}$ is adjacent to $v_{_{1}}$ in $G/(A-e)$. }
\label{MISaugmentable2}
\end{figure}

{\bf Proof.}
Take $\kappa$ to be the identity function. Let $Y'$ be a sub-solution. Then, by Lemma \ref{MisIndependentSolution}, $Y'$ is an independent set of vertices in $G$.   If in $G$  there is a vertex $v$   that is not adjacent to any vertex of $Y'$,  then $Y' \cup v$ is also a independent set of vertices in $G$. And by Lemma \ref{MisIndependentSolution}, $Y'\cup v$ is a feasible set.  Thus, $Y'$ is augmentable.  If there is no such a vertex $v$, then for all  vertices $v$ of $G$ that are not in $Y'$, $v$ is adjacent to some vertex of   $Y'$ in $G$. Hence, $Y'$ is a maximal independent set in $G$, as illustrated in Figure \ref{MISaugmentable2}. Thus $Y'$ is not a sub-solution. This is a contradiction.                                                
\qed\\

\begin{figure}[h]
\includegraphics[scale=0.42]{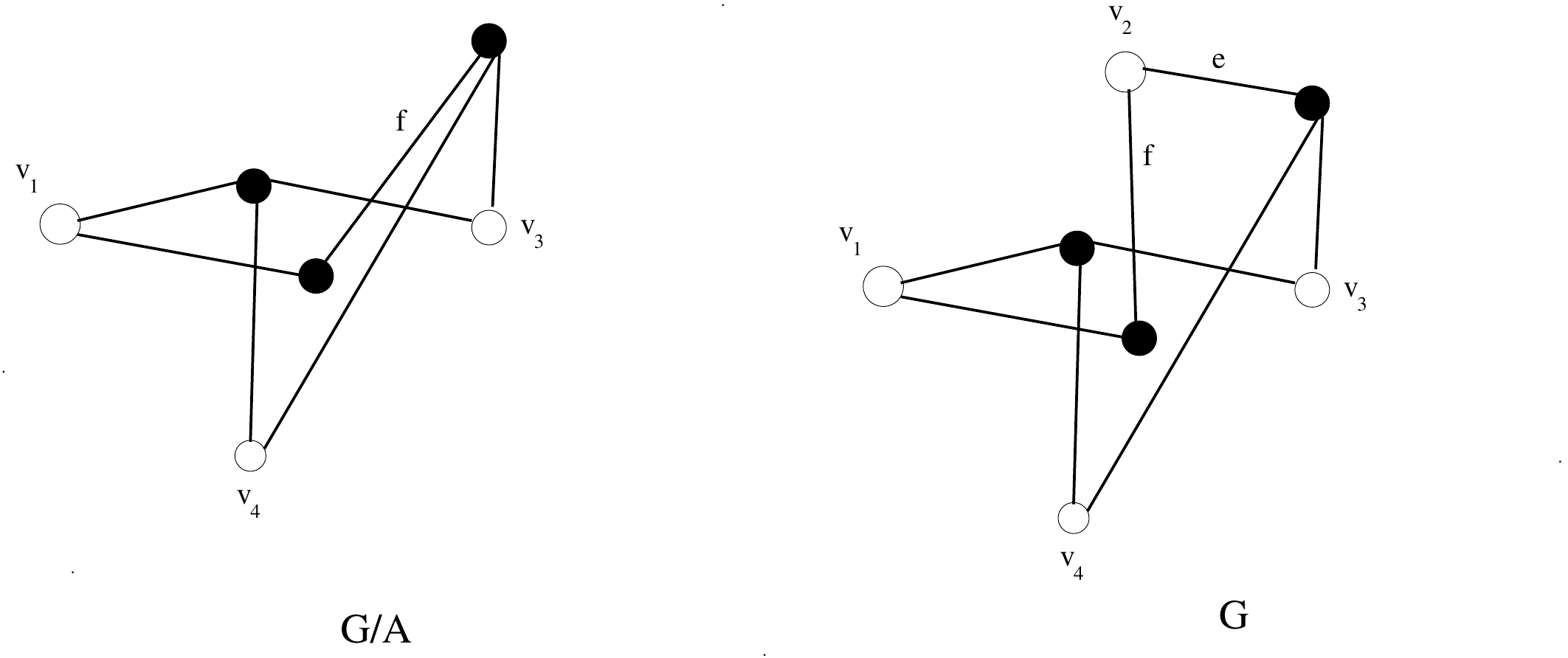}
\caption{The set $\{ v_{_{1}},  v_{_{3}}, v_{_{4}}\} $ is a maximal independent set of the graph $G/e= G/A$. The vertex $v_{_{2}}$ can be added to it to   obtain  the independent set $\{ v_{_{1}}, v_{_{2}}, v_{_{3}}, v_{_{4}}\} $ by re-inserting the edge $e$. }
\label{MISaugmentable1}
\end{figure}   

\begin{lemma}\label{MISAccessible} 
Every subset of a feasible set   of the Maximal Independent Set Problem is a feasible set.  That is,  the MISP  satisfies G1. That is, the set system $(X,\mathcal{I})$ of MISP is a simplicial complex (This seems to be  a feature of all P-complete and NP-complete  problems).
\end{lemma}

{\bf Proof.}
Let $Y'= \{v_{_{1}}, v_{_{2}}, \cdots , v_{_{k}} \}$ be a feasible set.  We aim to show that,  for any  vertex $v_{_{i}} \in \{v_{_{1}}, v_{_{2}}, \cdots , v_{_{k}} \}$, the set $Y' - v_{_{i}}$ is a feasible set.\\

So, let $Y'= \{v_{_{1}}, v_{_{2}}, \cdots , v_{_{k}} \}$ be a feasible set.  Then, there is a contraction-minor $G/A$, with $A \supseteq \emptyset$, such that $Y'$ is a maximal independent set in  $G/A$.
Consider any vertex  $ v_{_{i}}$ that is an element of $Y'$.  Since $G/A$ has no isthmus,   the vertex $v_{_{i}}$ is incident to an edge $e= (v_{_{i}}, w)$  such that  $ w \not \in Y'$. 

\begin{figure}[h]
\center
\includegraphics[scale=0.42]{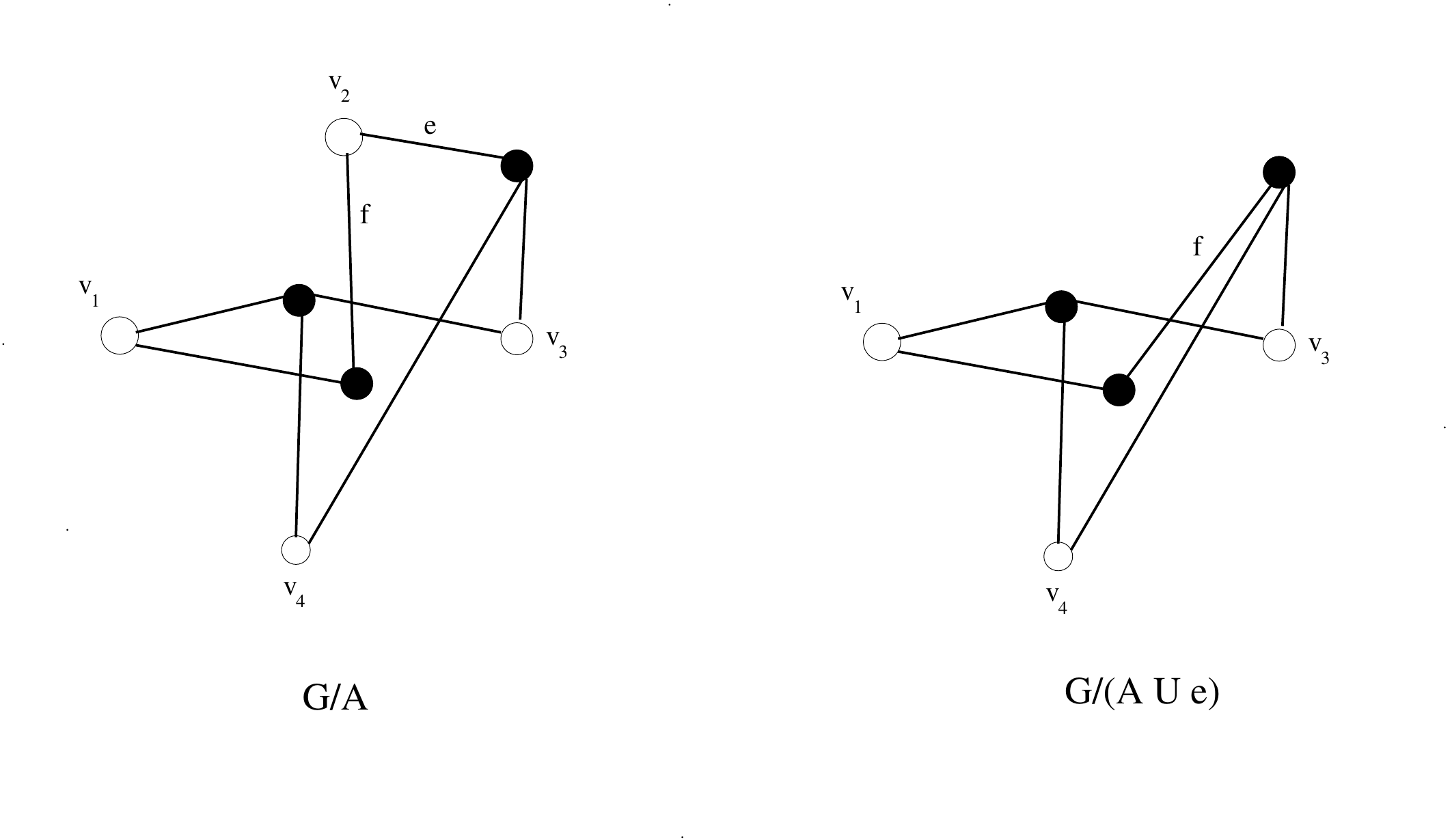}
\caption{The maximal independent set $\{ v_{_{1}},  v_{_{3}}, v_{_{4}}\} $  obtained from the independent set $\{ v_{_{1}}, v_{_{2}}, v_{_{3}}, v_{_{4}}\} $  by the contraction of the edge $e$. }
\label{MISaugmentable}
\end{figure}

(1)  If there is another edge $f= (w, v_{_{j}})$  such that $v_{_{j}}    \in Y'$, or   if there is another edge $f= (u, v_{_{i}})$  such that $u  \not   \in Y'$,  then contracting by $e$ yields the independent set   $Y' - v_{_{i}}$, which is  maximal  since any other vertex in $G/(A \cup e)$ but not in $Y' -v_{_{i}}$ is adjacent to some vertex in  $Y' - v_{_{i}}$, as illustrated in Figure \ref{MISaugmentable}. 

(2) If there is no such edge $f$, then there is a path $(v_{_{i}}- e_{_{1}}-v_{_{a}}-e_{_{2}}- v_{_{b}}- e_{_{3}}-v_{_{j}})$,  with $v_{_{i}}, v_{_{j}} \in Y'$,  and  at most two vertices $v_{_{a}}$ and $v_{_{b}}$ such that $v_{_{a}}$ and $v_{_{b}}$ are not elements of $Y'$, since $Y'$ is maximal in $G/A$.  By contracting by $e_{_{2}}=(v_{_{a}}, v_{_{b}})$  we get the graph $G/(A \cup e_{_{2}})$, where  $Y'$ is still a maximal independent set. And then  one falls in Case (1).

\qed\\

One may also check the following, which we do not prove formally since  we do not use it in the proof of the  main Theorem.  

\begin{lemma}\label{STPAccessible} 
Every subset of a  feasible set of the Spanning Tree Problem is a feasible set. That is, STP   problem is a simplicial complex.
\end{lemma}

\vspace{3mm}

\begin{lemma}\label{HamiltonianAccessibility}
Every  subset of a feasible set of the Hamiltonian Cycle Problem is a feasible set.  That is, the set system $(X,\mathcal{I})$ of HCP is   a simplicial complex.
\end{lemma}

{\bf Proof.}
Let $C= \{e_{_{1}}, e_{_{2}}, \cdots , e_{_{k}} \}$ be a hamiltonian cycle of the graph $G/A$, with $A \supseteq \emptyset$. Consider the  contraction of  the edge  $e_{_{r}} \in C$.   The contraction of the edge  $e_{_{r}}$ yields  the  cycle $C -  e_{_{r}} $, which is a hamiltonian cycle of the graph  $G/(A \cup e) $.    
\qed\\

\begin{figure}[h]
\center
\includegraphics[scale=0.42]{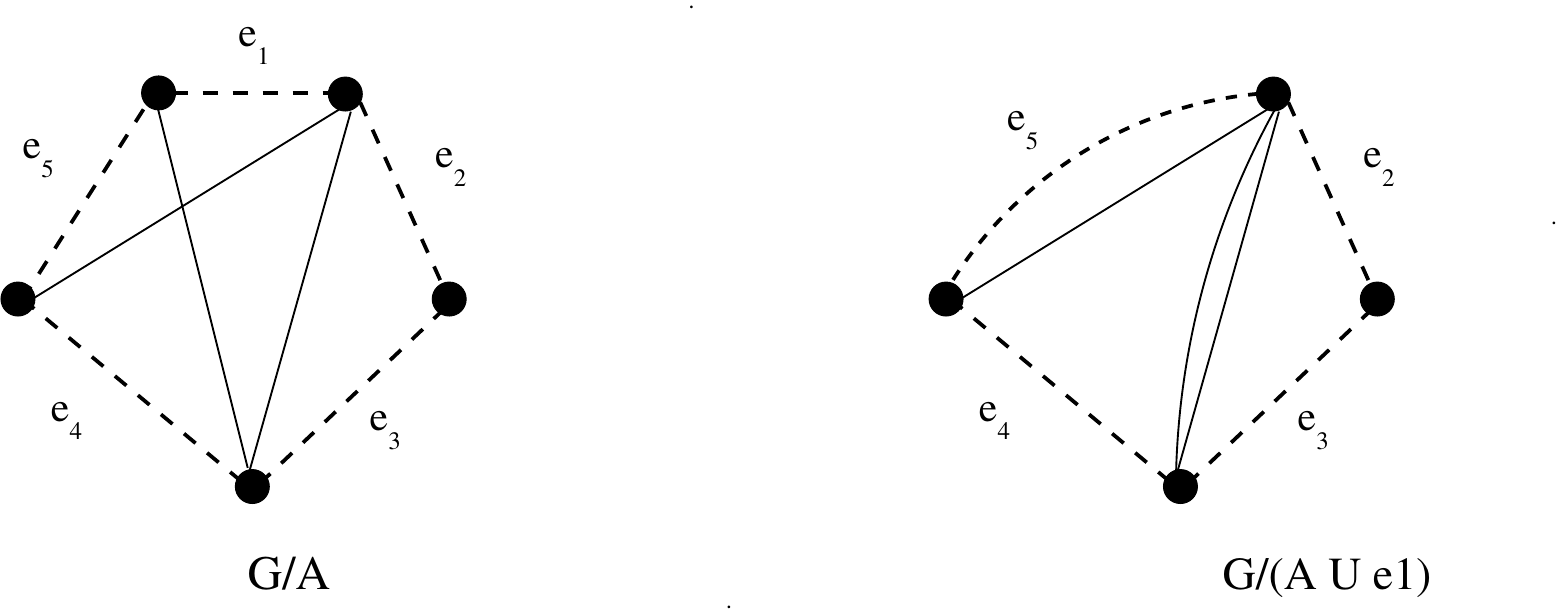}
\caption{The hamiltonian cycle $\{ e_{_{2}},  e_{_{3}}, e_{_{4}}, e_{_{5}}\} $  obtained from the hamiltonian cycle $\{ e_{_{1}}, e_{_{2}}, e_{_{3}}, e_{_{4}}, e_{_{5}}\} $  by the contraction of the edge $e_{_{1}}$. }
\label{HamiltonianAccessible}
\end{figure}   

\vspace{3mm}

 Most importantly for what follows in Corollary \ref{pnotnp},  we have the following important  observation.

\begin{lemma}\label{HCPnotAugmentable}
The Hamiltonian Cycle Problem is not fast-augmentable.
\end{lemma}

{\bf Proof.}


\begin{figure}[h]
\center
\includegraphics[scale=0.3]{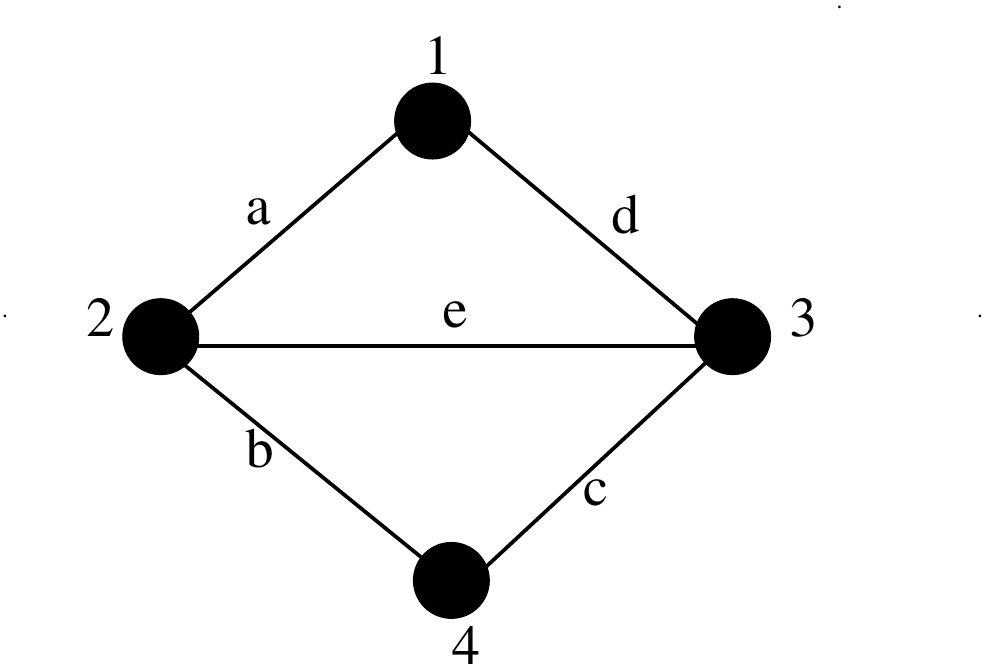}
\caption{A graph where a Hamiltonian circuit exists.}
\label{hamiltonian}
\end{figure}   

Consider   the graph  $G$ of Figure \ref{hamiltonian}. Let $X=E$, the set of edges of $G$. Let ${\Pi(G[X], \gamma)}$, denoted HCP, be the problem that consists of finding a Hamiltonian cycle of $G$. First, we would like to recall that, for the HCP problem on a graph $G$,  $Y'$ is a feasible set  means that $Y'$ is a Hamiltonian cycle of a contraction-minor $H$  of the graph $G$. Thus, the set of all the feasible sets is

\begin{eqnarray*}
\mathcal{I} &=& \{\emptyset, \textrm{all the singletons},\textrm{all the 2-subsets}, \{a,d,e\}, \{b,c,e\},\\
 && \{a,b,c\}, \{a,b,d\}, \{b,c,d\},\{a,c,d\}, \{a,b,c,d\}\}.
\end{eqnarray*}
 

 The subset $Y'= \{a,d,e\}$  is a feasible set (sub-solution), since it is a Hamiltonian cycle of the sub-instance $G/ \{b\}$. But there is no edge $x$ such that $x \in X - Y' $ and $Y' \cup x \in \mathcal{I}$.  That is, there is no contraction-minors of $G$ whose Hamiltonian cycle would be $Y' \cup x$, for all $x \in X - Y' $. Hence $Y'$ is not fast-augmentable. 
 
 \qed \\
 
 Indeed, this is a feature of all the feasible sets that are cycles in the graph $G$.
 
 \begin{lemma}\label{CycleNotAugmentable}
 Consider the HCP on the  connected  isthmus-less graph $G$. If $C'$ is a  non-Hamiltonian cycle in $G$, then $C'$ is a feasible set of HCP that is not fast-augmentable.  And there is no Hamiltonian cycle  $C$ such that $C' \subset C$.
 \end{lemma}
 
 {\bf Proof.}
  $C'$ is a feasible set since it is a Hamiltonian cycle of the graph $G/D$, where  $G/D$ is defined as follows: if there is a cycle  $C_{_{1}} \not = C'$,  contract all the edges in  $C_{_{1}}-C'$ except one of them. Repeat recursively these operations of contractions.\\
 
 Now, if $C$ is a hamiltonian cycle of $G$, then $C$   must be of the form $C= (C'-A)\cup B$, where $A \subset C'$. Hence $C'$ can not be fast-augmentable, and $C' \not \subset C$.  
 
 \qed
 
  Notice that in Figure \ref{hamiltonian}  the set  $\{a,d,c\}$  is  a feasible set  that  is a Hamiltonian cycle of the sub-instance $G/ \{b\}$ as well. And  $\{a,d,c\}$  is fast-augmentable.  On the other hand,   we also  have that the Maximum Matching  Problem (MMP) is not fast-augmentable, and yet,  it is solvable in polynomial time.  Indeed,  if $Y'$ is a non-fast-augmentable feasible set  of the MMP Problem $\Pi(G[X],\gamma)$, then, as shown by  the  `Augmenting Path algorithm'  from Edmonds \cite{edmonds1}, there is a polynomial-time function $\kappa$ that  transforms $Y'$ into another sub-solution $Z'$, where $Z'$ is fast-augmentable.  Thus,  MMP   satisfies the Augmentability Axiom.  Therefore,    we  have to show  that the Hamiltonian Cycle Problem  can not satisfy the Augmentability Axiom by showing that there may not be a polynomial time function $\kappa$ that transforms its non-fast augmentable feasible set $Y'=  \{a,d,e\}$  into a fast-augmentable feasible set $Z'=  \{a,d,c\}$. The following definitions are instrumental to that end.\\

{\bf Definitions  2.4.}
Consider the  accessible  problem      $\Pi(G[X], \gamma)$.  
We say that  $\Pi(G[X], \gamma)$ is  \textit{fast-accessible} if given  any feasible set $I$, there is an  accessibility chain $\emptyset \unlhd I^{^{(1)}} \unlhd I^{^{(2)}} \unlhd  \cdots  \unlhd I$ such that, for $1 \leq r \leq |I|$,  every transition from  $   I^{^{(r)}}$  down to  $I^{^{(r-1)}} $ is done via a  single contraction,  of an   edge $e$, say.  The problem $\Pi(G[X], \gamma)$  is \textit{slow-accessible} if there is an instance $G[X]$ such that in all the accessibility chains $\emptyset \unlhd I^{^{(1)}} \unlhd I^{^{(2)}} \unlhd  \cdots  \unlhd I$, some transitions  from  $   I^{^{(r)}}$  down to  $I^{^{(r-1)}}$  require  at least two contractions, of edges $e$ and $f$, say. \\
  
  An illustration  of  a problem that is slow-accessible  is the Maximum Matching Problem,  as  shown in Figure  \ref{Matching2}  below.  This example is also instrumental in understanding  the proof  of Lemma \ref{kappaEntailsSlowAccessible}, where  we show that \textit{slow-accessibility}  is a feature  of all the polynomial solvable problems where there exists  such a function  $\kappa$  that transforms in polynomial time a  non-fast augmentable feasible set   into a fast-augmentable feasible set.    However, for the HCP Problem, we make the following observation.  \\

 \begin{figure}[H]
 \center
 \includegraphics[scale=0.52]{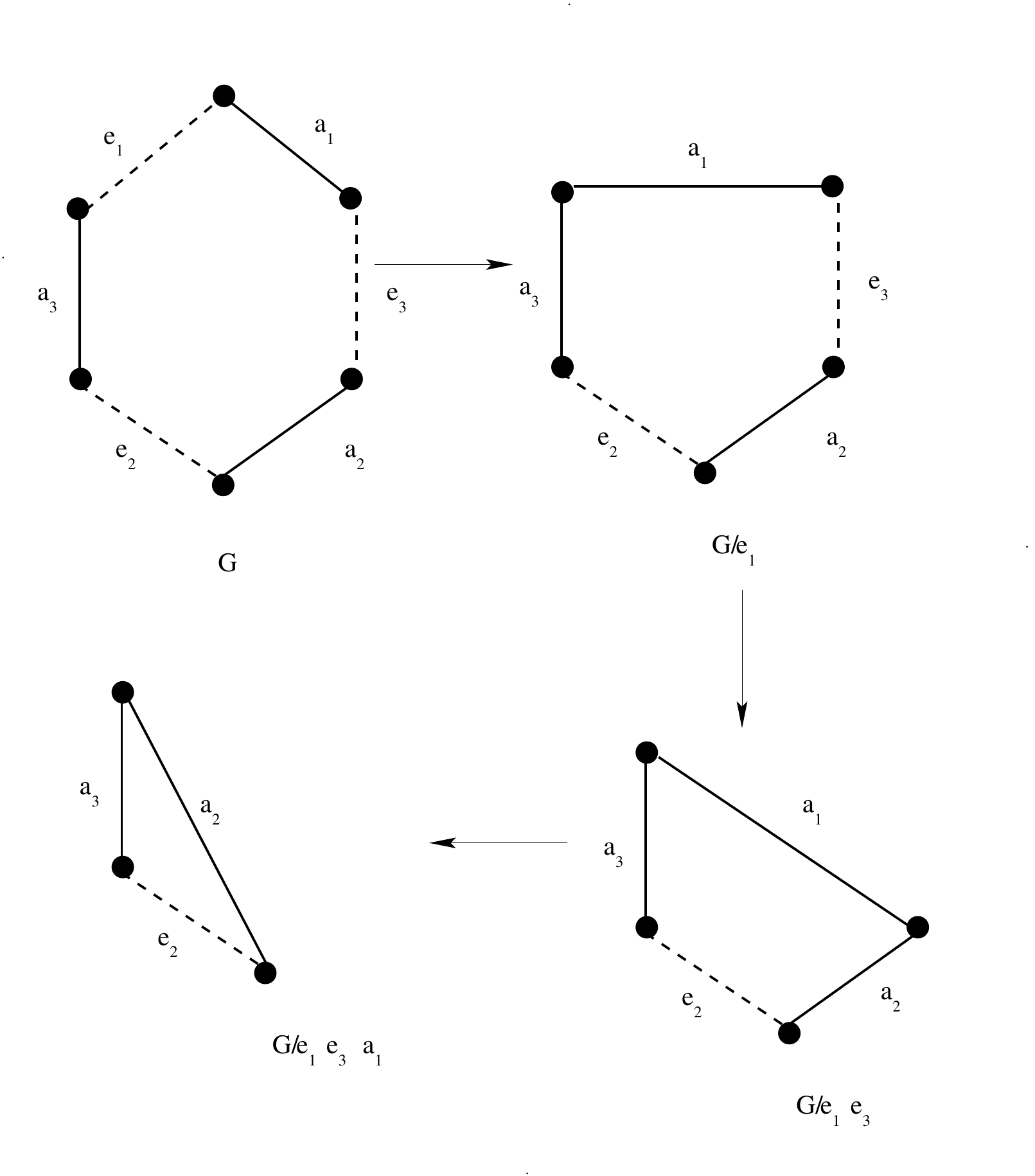}
 \caption{ $E= \{ e_{_{1}}, e_{_{2}}, e_{_{3}}\}$ and   $A= \{ a_{_{1}}, a_{_{2}}, a_{_{3}}\}$ are both maximum matchings in the graph $G$.  Now,  contracting by the edge $e_{_{1}}$ automatically  removes the edge $a_{_{1}}$ from the maximum matching $A$. Hence both $E-e_{_{1}}$ and $A-a_{_{1}}$ are maximum matching in the graph $G/e_{_{1}}$.  But  Contracting  $G/e_{_{1}}$ by the edge $e_{_{3}}$  does not   remove the edge $a_{_{2}}$ or  $a_{_{3}}$ from the  maximum matching   $A-a_{_{1}}$ .  Thus, $E-e_{_{1}}-e_{_{3}}$ is not a maximum matching in  the graph $G/e_{_{1}}/e_{_{3}}$, but   $A-a_{_{1}}$ still is.  One needs to contract by the edge $a_{_{1}}$  to make  $E-e_{_{1}}-e_{_{3}}$  a maximum matching in  $G/e_{_{1}}/e_{_{3}}/a_{_{3}}$. Hence the MMP Problem is slow-accessible.  } 
 \label{Matching2}
 \end{figure}

\begin{lemma}\label{HCPnotSlowAccessible}
The Hamiltonian Cycle Problem is fast-accessible
\end{lemma}

{\bf Proof.}
Let $Y$ be a Hamiltonian cycle of  a graph $H$. Then $Y-e$ is a Hamiltonian cycle of $H/e$ for every $e \in Y$.
\qed\\

 From Lemma \ref{MISAugmentable}  to  Lemma \ref{HCPnotSlowAccessible}, we have identified clear combinatorial  differences between  a problem that is in the computational class P-complete and  a problem that is in the computational class NP-complete.  Indeed, MISP   satisfies Axioms G1 and M2' while HCP  satisfies G1 but not M2'. If only we show that all problems solvable in polynomial time must obey Axiom M2', then we have shown that  the computational class $P$ is different from  the computational class $NP$.\\

 \vspace{3mm}

The  proof of Theorem \ref{MainTheorem} requires Lemma  \ref{claim 0},   which shows that, given a Graphical Search Problem Solvable in Polynomial Time $\Pi(G[X],\gamma)$    and    an ordering  by minor-contractions  of the sub-instances of the  problem $\Pi(G[X],\gamma)$,  the functions $\phi$  and $\psi$ induce an ordering   of the sub-instances of the MISP,  and vice-versa, as  shown schematically  in Figures \ref{figureParallelChains1}.

\begin{figure}[h]
\center
\begin{tikzpicture}[scale=1.6]
\draw [->>]  (5,0.9)-- (5,0.1);
\node [right] at (5,0.5)%
           {$\phi$};
\node[align=left, below] at (5.2,0)%
    {$\phi(G)$};
\node[align=left, above] at (5,1)%
    {G};    
\node[align=left, below] at (0,0)%
    {$\emptyset$};
\node[align=left, above] at (0,1)%
    {$\emptyset$}; 
 \draw [->>]  (0,0.9)-- (0,0.1); 
\node [left] at (0,0.5)%
           {$\phi$};      
 

 \node[ above] at (2.7,1)%
    {$G_{_{2}}$};     
   \node[ below] at (2.7,0)%
    {$\phi(G_{_{2}})$};     
\draw [->>]  (2.7,0.9)-- (2.7,0.1); 
\node [left] at (2.7,0.5)%
           {$\phi$};       
         
\node[align=left, below] at (1.7,0)%
    {$\phi(G_{_{1}})$};
\node[align=left, above] at (1.7,1)%
    {$G_{_{1}}$};  
\draw [->>]  (1.7,0.9)-- (1.7,0.1); 
\node [left] at (1.7,0.5)%
           {$\phi$}; 
           
 \node[align=left, below] at (2.2,0)%
    {$\prec$};
\node[align=left, above] at (2.2,1)%
    {$\prec$};            
               
 \node[align=left, below] at (1.2,0)%
    {$\prec$};
\node[align=left, above] at (1.2,1)%
    {$\prec$}; 
    
 \node[align=left, below] at (0.2,0)%
    {$\prec$};
\node[align=left, above] at (0.2,1)%
    {$\prec$}; 
    
 \node[align=left, below] at (3.2,0)%
    {$\prec$};
\node[align=left, above] at (3.2,1)%
    {$\prec$};  
    
 \node[align=left, below] at (4.7,0)%
    {$\prec$};
\node[align=left, above] at (4.7,1)%
    {$\prec$};
    
  \node[align=left, below] at (4,0)%
    {$\cdots$};
\node[align=left, above] at (4,1)%
    {$\cdots$};   
    
 \node[align=left, below] at (0.9,0)%
    {$\cdots$};
\node[align=left, above] at (0.9,1)%
    {$\cdots$};                                   
 
\end{tikzpicture}
\caption{ Let $G$  denote  $G[X]$, $G_{_{1}}$ denote $G[X_{_{1}}]$, $G_{_{2}}$ denote $G[X_{_{2}}]$, and let $\prec$ denote `contraction-minor'.  `Parallel' chains of  contractions  from instances  of $\Pi(G[X],\gamma)$  to instances  of  MISP.}  
\label{figureParallelChains1}
\end{figure}

\begin{lemma}\label{claim 0}
Let $G[ X_{_{1}}]$ and $G[X_{_{2}}]$  be  contraction-minors  of $G[X]$.  For $i=1,2$,  let $Y_{_{i}}$ and $S_{_{i}}$ be solutions of  $\Pi(G[X_{_{i}}],\gamma)$ and   $\phi(\Pi(G[X_{_{i}}],\gamma))$, respectively. Then,

\begin{enumerate}
\item $ G[X_{_{1}}]$  is a contraction-minor of $ G[X_{_{2}}]$  if and only $ \phi(G[X_{_{1}}])$  is a contraction-minor of $ \phi(G[X_{_{2}}])$.
\item      $Y_{_{1}} \subset Y_{_{2}}$ if and only if      $S_{_{1}} \subset S_{_{2}}$.

\end{enumerate}
\end{lemma}

{\bf Proof}\\

 \begin{enumerate}
 
\item      Consider the problem   $\Pi(G[X_{_{2}}], \gamma)$.  The graph  $G[X_{_{1}}]$ is a contraction-minor of  $G[X_{_{2}}]$ if and only if   $Y_{_{1}}$  is a sub-solution of the problem  $\Pi(G[X_{_{2}}], \gamma)$ by Definition 2.2.  Now $Y_{_{1}}=\psi(S_{_{1}})$. Hence,   $Y_{_{1}}$  is a sub-solution of the problem  $\Pi(G[X_{_{2}}], \gamma)$ if and only if     $S_{_{1}}$ is a sub-solution of the problem MISP instanced on the input  $\phi(G[X_{_{2}}])$.   Finally, by Definition 2.2, we have that  $S_{_{1}}$ is a sub-solution of the  MISP instanced on the input  $\phi(G[X_{_{2}}])$ if and only if  $\phi(G[X_{_{1}}])$ is a contraction-minor of  $\phi(G[X_{_{2}}])$.\\  

 \vspace{3mm}

    \item By Lemma \ref{MISAccessible} ,  $S_{_{1}} \subset S_{_{2}}$   $\Longleftrightarrow$ $\phi(G[X_{_{1}}])$ is a contraction-minor of $\phi(G[X_{_{2}}])$ $\Longleftrightarrow$ $G[X_{_{1}}]$ is a contraction-minor of $G[X_{_{2}}]$ ,  by Part (1).  Finally  $G[X_{_{1}}]$ is a contraction-minor of $G[X_{_{2}}] \implies Y_{_{1}} \subset Y_{_{2}}$.\\

    Now, to prove that $Y_{_{1}} \subset Y_{_{2}} \implies S_{_{1}} \subset S_{_{2}}$, suppose that  $Y_{_{1}} \subset Y_{_{2}}$  but      $G[X_{_{1}}]$ is not  a contraction-minor of $G[X_{_{2}}]$.  Since $G[X_{_{1}}]$ is a contraction-minor of $G[X]$,  then, there is a sub-instance $G[X_{_{3}}]$,  with a solution $Y_{_{3}}$, such that    $G[X_{_{1}}]$ is a contraction-minor of  $G[X_{_{3}}]$ and, thus, $Y_{_{1}} \subseteq Y_{_{3}}$.  (Note that  $X_{_{3}}$ may be $X$).   If  $G[X_{_{3}}]  = G[X_{_{2}}]$, then we are done.  \\

    Now suppose that $G[X_{_{3}}] \not = G[X_{_{2}}]$. We recall that  $\phi(G[X_{_{1}}])$ may contain many maximal independent sets of vertices, and some of them may not be subset of $S_{_{2}}$. We only have to show there is at least one such  maximal independent sets of vertices which is a subset of $S_{_{2}}$. \\

   Let $R_{_{1}}= \{ v_{_{1}}, \cdots, v_{_{m}}\}$ be any  maximal independent set of vertices (a solution of $\phi(G[X_{_{1}}])$.   If  the vertex  $v$ belongs to the feasible set  $R_{_{1}}$, then there is an element   $a \in  Y_{_{1}}$  such that $a= \psi(v)$. Now, since   $Y_{_{1}} \subset Y_{_{2}}$, and  $Y_{_{1}} \subset Y_{_{3}}$,  we have that $ a \in Y_{_{2}}$ and  $ a \in Y_{_{3}}$.  Hence   $a= \psi(w) $ for a vertex $w \in S_{_{2}}$ and  $a= \psi(v) $ for the vertex $v \in S_{_{3}}$, since  $R_{_{1}} \subset S_{_{3}}$ by Part (1).   Now, suppose that $v \in R_{_{1}}$, but $v\not \in S_{_{2}}$ for some $v \in R_{_{1}}$, then $S_{_{3}} \not = S_{_{2}}$.  Since  for every such vertex $v_{_{k}}$  such that  $v_{_{k}} \in S_{_{3}}$ but   $v_{_{k}} \not \in S_{_{2}}$, we have that    $a_{_{k}}= \psi(w_{_{k}}) $ for a vertex $w_{_{k}} \in S_{_{2}}$ and  $a_{_{k}}= \psi(v_{_{k}}) $ for the vertex $v_{_{k}} \in S_{_{3}}$,  let $S_{_{2}}= \{ w_{_{1}}, \cdots, w_{_{m}}, b_{_{1}}, \cdots b_{_{l}}\}$  and   $S_{_{3}}= \{ v_{_{1}}, \cdots, v_{_{m}}, c_{_{1}}, \cdots c_{_{r}}\}$, where some $w$, but not all of them, may be equal to some $v$. And,  since both $S_{_{2}}$ and $S_{_{3}}$ are   independent sets of vertices, we have that  $S'_{_{2}}= \{ w_{_{1}}, \cdots, w_{_{m}}\}$  and   $S_{_{3}}'= \{ v_{_{1}}, \cdots, v_{_{m}}\}$ are also independent sets of vertices. Thus, by Lemma \ref{MISAccessible} , there  exist graphs $T_{_{2}}$ and $T_{_{3}}$, that are   contraction-minors of  $\phi(G[X_{_{2}}])$ and $\phi(G[X_{_{3}}])$, respectively,   such that $S_{_{2}}'$ and $S_{_{3}}'$ are maximal  independent sets of vertices in $T_{_{2}}$ and $T_{_{3}}$, respectively.  However, $S_{_{3}}'= R_{_{1}}$.  Hence $\psi(S_{_{3}}')=  \psi(R_{_{1}})=  Y_{_{1}}$. Moreover,  $\psi(S_{_{2}}')= Y_{_{1}}$,  since  $\psi $ maps each  element $v_{_{k}}$ of $S_{_{3}}'$ and each element $w_{_{k}}$  of $S_{_{2}}'$ onto the same element $a_{_{k}}$ of $Y_{_{1}}$.   Therefore $T_{_{3}} = T_{_{2}}$,   lest   $\phi$ maps $G[X_{_{1}}]$ to two different instances of MISP.  Thus  $T_{_{3}} = T_{_{2}}$ is a contraction-minor of   $\phi(G[X_{_{2}}])$ that contains two different solutions  $S_{_{2}}'$ and $S_{_{3}}'$, where  $S_{_{2}}' \subset S_{_{2}}$.  Now,  since  $\psi(S_{_{2}}')= Y_{_{1}}$, we get that there is a subset of $S_{_{2}}$ that is a solution of   $\phi(G[X_{_{1}}])$.\\ \\

  \end{enumerate}

\qed\\

\vspace{3mm}

\section{  Main Theorem and Proof} 


\vspace{3mm}

 \vspace{3mm}

  \begin{theo}\label{MainTheorem}
  Let $G$ be an isthmus-less connected labelled graph with vertex-set $V$ and edge-set $E$, and let $X$ be either $V$ or $E$.   Let $ \mathcal{I}$ be  the set of all the feasible sets    of  the accessible search problem $\Pi(G[X], \gamma)$. 
  The  problem $ \Pi(G[X],\gamma)$    is solvable in polynomial time if and only if, for every input $G[X]$,  the set system $(X, \mathcal{I})$   satisfies Axiom M2'. That is, 
all its non-basic feasible sets  (sub-solutions)  are augmentable.
\end{theo} 

{\bf Proof.}

The  proof uses the facts that the search  problem  MISP  satisfies  Axiom  $M2'$ and  $\hat {MISP}$, the decision problem associated to MISP,    is  P-complete.  Hence, given an instance  $\hat \Pi(G[X], \gamma)$, there is an algorithm  $\phi$ that transforms  $\hat \Pi(G[X], \gamma)$ into an instance of  $\hat {MISP}$ and an algorithm  $\psi$  that  transforms each solution of $\hat MISP$  into a solution of $\hat \Pi(G[X], \gamma)$, such that $S$ is a solution  $\hat {MISP}$ if and only if $\psi(S)$ is a solution of $\hat \Pi$.  Thus, if $Y'$ is a sub-solution of a polynomial time search problem $\Pi(G[X], \gamma)$, there is a  contraction-minor of $G[X]$, denoted   $G[X']$,  such that $Y'$ is a solution of the problem $\hat \Pi(G[X'], \gamma)$. Hence, there is a sub-solution $S'$ of MISP, such that $Y'= \psi(S')$ and $S'$ is a solution of $\phi(G[X'])$.   Now since $S'$ is augmentable as $S'\cup v$, the   proof  aims to show that $\psi(S'\cup v)$ is an augmentation of $\kappa(Y')$, where $\kappa$ is another polynomial time computable function from the set of feasible sets of $\Pi(G[X], \gamma)$ to the set of feasible sets of $\Pi(G[X], \gamma)$.  That is,  every  polynomial time search problem satisfies $M2'$.  Conversely, if every sub-solution $Y'$ is augmentable and every solution $Y$  is accessible, then there is a steady path   $\emptyset \unlhd Y^{^{(1)}} \unlhd Y^{^{(2)}} \unlhd  \cdots  \unlhd Y$. That is, the solution $Y$ can be found in polynomial time, by using a `Generalised Greedy Algorithm'.   \\


\begin{enumerate}
\item{\bf Necessity.}

\vspace{3mm}

 By Theorem \ref{maximalISEasy}, we have that $\hat {MISP}$  Problem is in  $\mathcal{P}$-complete. And,  by Lemma \ref{MISAugmentable}, we have that  the search problem associated with $\hat {MISP}$  satisfies Axiom M2'. \\

Now, let $\Pi(G[X],\gamma)$  be a  Graphical Search Problem solvable in polynomial time with a solution $Y$.  We  aim to show that  the set system    $(X,\mathcal{I})$ satisfies Axiom M2'.\\

  Indeed,  if $\Pi(G[X],\gamma)$  is a  Graphical Search Problem solvable in polynomial time, then the decision problem  $\hat \Pi(G[X], \gamma)$ is in $\mathcal{P}$,  and, by Definition \ref{reducibility} and Theorem \ref{maximalISEasy}, there is an algorithm $\phi$ that transforms   the instance $G[X]$ of $\hat \Pi$ into an  instance  $\phi(G[X])$ of $\hat {MISP}$, and  there is an algorithm $\psi$  that  transforms a solution  $S$ of $\hat {MISP}$  into the solution $Y$ of $\hat \Pi$ such that $S$ is a solution of $\hat {MISP}$ if and only if $\psi(S)=Y$ is a solution of $\hat \Pi$. \\

  Let $Y'$ be a sub-solution  of $\Pi(G[X],\gamma)$. That is, $Y'$ is a solution   of the search problem  $\Pi(G[X'], \gamma)$, where $X' \subset X$ and $G[X']$ is a contraction-minor of $G[X]$.  Consider the feasible set $S'$ of MISP which is a solution of the MISP instance  $\phi(G[X'])$, and $S'$ is augmentable as  $S'\cup v$. That is, $\psi(S')=Y'$.  We aim at showing that either (A): if there is a solution $Y$ such that $Y' \subset Y$, then, either    there is an accessibility chain $\emptyset \unlhd \cdots \unlhd Y' \unlhd Y'\cup x \unlhd \cdots \unlhd Y$, such that $\psi(S')=Y$' and $Y' \cup x= \psi (S'\cup v)$ is an augmentation of $Y'$,  as illustrated in in Figure \ref{figureParallelChains}, or there is a polynomial time computable function $\kappa$ that transforms $Y'$  into another sub-solution $Z'$   such that $Z' \cup z= \psi(S'\cup v )$. Or (B): if there is no solution $Y$ such that $Y' \subset Y$, then there is a polynomial time computable function $\kappa$ that transforms $Y'$  into another sub-solution $Z'$   such that $Z' \cup z= \psi(S'\cup v )$. (Please note that $\kappa$ is just a generalisation of Edmond's Augmenting Path Algorithm for the Maximal  Matching Problem. \cite{edmonds1} ).  \\

(A)  Suppose  there is a solution $Y$ such that $Y' \subset Y$.   Since $Y$ is accessible,  there is an accessibility chain from $\emptyset$  to $Y$. \\
\\

 First, we have that $S'$ such that $Y'= \psi(S')$  is a sub-solution (not a basis). Indeed, if it is a basis, then $\phi(G[X])= G[cl(S')]$ and $\phi(G[cl(Y')])= G[cl(S')]$. Thus $\psi(S')= Y'$, and  $\psi(S')= Y$.   But, since $Y'$ is a sub-solution, $Y' \not =Y$.   Hence,   $\psi$ is not well defined. This is a contradiction.\\

  Now, since $S'$ is a sub-solution, by Lemma \ref{MISAugmentable}, $S'$ is augmentable. That is, there is a vertex $v \in V( \phi(G[X])) - S'$ such that $S'\cup v$ is a feasible set. Consider the accessibility chain $\emptyset \unlhd \cdots \unlhd S' \unlhd S'\cup v \unlhd \cdots \unlhd S$, where $S$ is a solution of $\phi(G[X])$. This chain exists, since MIS satisfies M1 and M2'.   By Lemma  \ref{claim 0},  we have  that  such a  chain  induces a parallel chain on the set of subsets of $Y$. Suppose that  there is an accessibility chain  containing $Y'$, as illustrated  in Figure \ref{figureParallelChains}.  Then, $Y' \cup x$ is a feasible set, and $Y'$ is fast-augmentable.

\begin{figure}[h]
\center
\begin{tikzpicture}[scale=1.6]
\draw [->>]  (5,0.9)-- (5,0.1);
\node [right] at (5,0.5)%
           {$\psi$};
\node[align=left, below] at (5,0)%
    {Y};
\node[align=left, above] at (5,1)%
    {S};    
\node[align=left, below] at (0,0)%
    {$\emptyset$};
\node[align=left, above] at (0,1)%
    {$\emptyset$}; 
 \draw [->>]  (0,0.9)-- (0,0.1); 
\node [left] at (0,0.5)%
           {$\psi$};      
 

 \node[ above] at (2.7,1)%
    {$S' \cup v$};     
   \node[ below] at (2.7,0)%
    {$Y' \cup x$};     
\draw [->>]  (2.7,0.9)-- (2.7,0.1); 
\node [left] at (2.7,0.5)%
           {$\psi$};       
         
\node[align=left, below] at (1.7,0)%
    {$Y'$};
\node[align=left, above] at (1.7,1)%
    {$S'$};  
\draw [->>]  (1.7,0.9)-- (1.7,0.1); 
\node [left] at (1.7,0.5)%
           {$\psi$}; 
           
 \node[align=left, below] at (2.05,0)%
    {$\unlhd$};
\node[align=left, above] at (2.05,1)%
    {$\unlhd$};            
               
 \node[align=left, below] at (1.4,0)%
    {$\unlhd$};
\node[align=left, above] at (1.4,1)%
    {$\unlhd$}; 
    
 \node[align=left, below] at (0.2,0)%
    {$\unlhd$};
\node[align=left, above] at (0.2,1)%
    {$\unlhd$}; 
    
 \node[align=left, below] at (3.2,0)%
    {$\unlhd$};
\node[align=left, above] at (3.2,1)%
    {$\unlhd$};  
    
 \node[align=left, below] at (4.7,0)%
    {$\unlhd$};
\node[align=left, above] at (4.7,1)%
    {$\unlhd$};
    
  \node[align=left, below] at (4,0)%
    {$\cdots$};
\node[align=left, above] at (4,1)%
    {$\cdots$};   
    
 \node[align=left, below] at (1,0)%
    {$\cdots$};
\node[align=left, above] at (1,1)%
    {$\cdots$};                                   
 
\end{tikzpicture}
\caption{`Parallel' accessibility chains of a solution $S$ of MIS and the solution $Y$ of $\Pi(X)$ such that $\psi(S)=Y$. }
\label{figureParallelChains}
\end{figure}

 Now,  suppose that $Y'$ is not fast-augmentable. That is, suppose that, for all $x \in X - Y'$, $Y' \cup x$ is not a feasible set. Then,  for all $x \in X - Y'$, $x$ belongs to some  $cl(Y')$.\\

 
 
 

 If $cl(Y')$ is unique, then $X - Y' \subseteq cl(Y')$. Thus  $cl(Y')=X$. Therefore, $\phi(G[X])=G[cl(S')]$. But, we also  have that $\phi(G[X])= G[cl(S)] $, and  since $cl(S') \subset cl(S' \cup v) \subseteq cl(S)$, we have  that $\phi$  maps $G[X]$  to two different  instances of $\hat{MISP}$. Thus $\phi$ is not a well-defined function. This is a contradiction.  Hence, $cl(Y')$ is not unique.\\
 
  Suppose there are $s$ different $cl(Y')$, each being of the form $Y' \cup P_{_{j}}$. 
 Now, for a fixed element $x_{_{1}} \in P_{_{1}}$ consider taking  the element  $x_{_{k}}$ such that  $x_{_{k}} \in P_{_{k}}$, $k \not =1$, and consider the graph   $G[P_{_{k}}\cup P_{_{1}} \cup Y']$. That is, the graph whose edge-set or vertex-set is the set $P_{_{k}}\cup P_{_{1}} \cup Y'$, and which is  constructed  by re-insertion from a contraction-minor $H$ such that $Y'$ is a solution of $\Pi(H, \gamma)$.  \\
 
  If $Y' \cup x_{_{k}}$ or $Y' \cup x_{_{1}}$ is a solution of the instance $G[P_{_{k}} \cup P_{_{1}} \cup Y']$  , then $Y'$ is augmentable. This is a contradiction. 
Therefore, for all $k$ in $2 \cdots s$, we have that    a solution of $G[P_{_{k}} \cup P_{_{1}} \cup Y']$ must be $Y' \cup x_{_{k}} \cup x_{_{i}} \cup \cdots \cup x_{_{j}}$, where $x_{_{i}}, \cdots , x_{_{j}} \in P_{_{1}}$ or $P_{_{k}}$ and $j > 1$. That is, a solution  of  $G[P_{_{k}} \cup P_{_{1}} \cup Y']$  must contain more than one element of $X - Y'$. Without loss of generality, suppose that it contains two elements, $x_{_{j}}$ and $x_{_{k}}$, where $x_{_{k}} \in P_{_{k}}$ and $x_{_{j}} \in P_{_{1}}$. Thus, suppose that  $Y' \cup x_{_{j}} \cup x_{_{k}}$ is a feasible set, but neither $Y' \cup x_{_{j}}$  nor $Y' \cup x_{_{k}}$  is a feasible set, and  suppose  that this holds for all the  $P_{_{k}}$. \\

 We have that $Y' \cup x_{_{1}} \cup x_{_{j}}$ is a feasible set. But  since  $Y' \cup x_{_{1}} \cup x_{_{j}}$ is accessible,  there is an element $x$ in  $Y' \cup x_{_{1}} \cup x_{_{j}}$ such that $A = (Y' \cup x_{_{1}} \cup x_{_{j}})- x$  is a feasible set. If $x=x_{_{1}}$ or $x=x_{_{j}}$  then either $Y' \cup x_{_{j}}$  or $Y' \cup x_{_{1}}$  must be a feasible set.  This is  a contradiction. Thus  $x \in Y'$, and  we have that $Y' \subset  Y' \cup x_{_{1}} \cup x_{_{j}}$ and $A \subset Y' \cup x_{_{1}} \cup x_{_{j}}$, and $A$ is augmentable since $A\cup x= Y$ . So define $\kappa(Y')=A$. \\

  We have that  the function $\kappa$ runs in polynomial time. Indeed,   given $Y'$,  finding $S'$ such that $\psi(S')= Y'$, and   augmenting $S'$ into $S'\cup v$, and  finding $Y= Y'\cup x_{_{1}} \cup x_{_{k}}= \psi(S'\cup v)$, and finding $x \in Y'$ such that $A=Y-x$ can be done in polynomial time. \\

(B)  There is no solution $Y$ such that $Y' \subset Y$.  Suppose $Y'$ is a solution of the problem $\Pi(H, \gamma)$, where $H$ is a contraction-minor of $G[X]$.     Since $\Pi(G[X], \gamma)$ is solvable in polynomial time,  we want to show, by construction, that  there is a polynomial algorithm  $\kappa$ that transforms $Y'$ into   $Z'$, where $Z'$ is another sub-solution of  $\Pi(H, \gamma)$, and $Z'$ is fast-augmentable  as $Z' \cup z$,  for some element $z \in X-Z'$. \\

\begin{figure}[h]
\center
\includegraphics[scale=0.5]{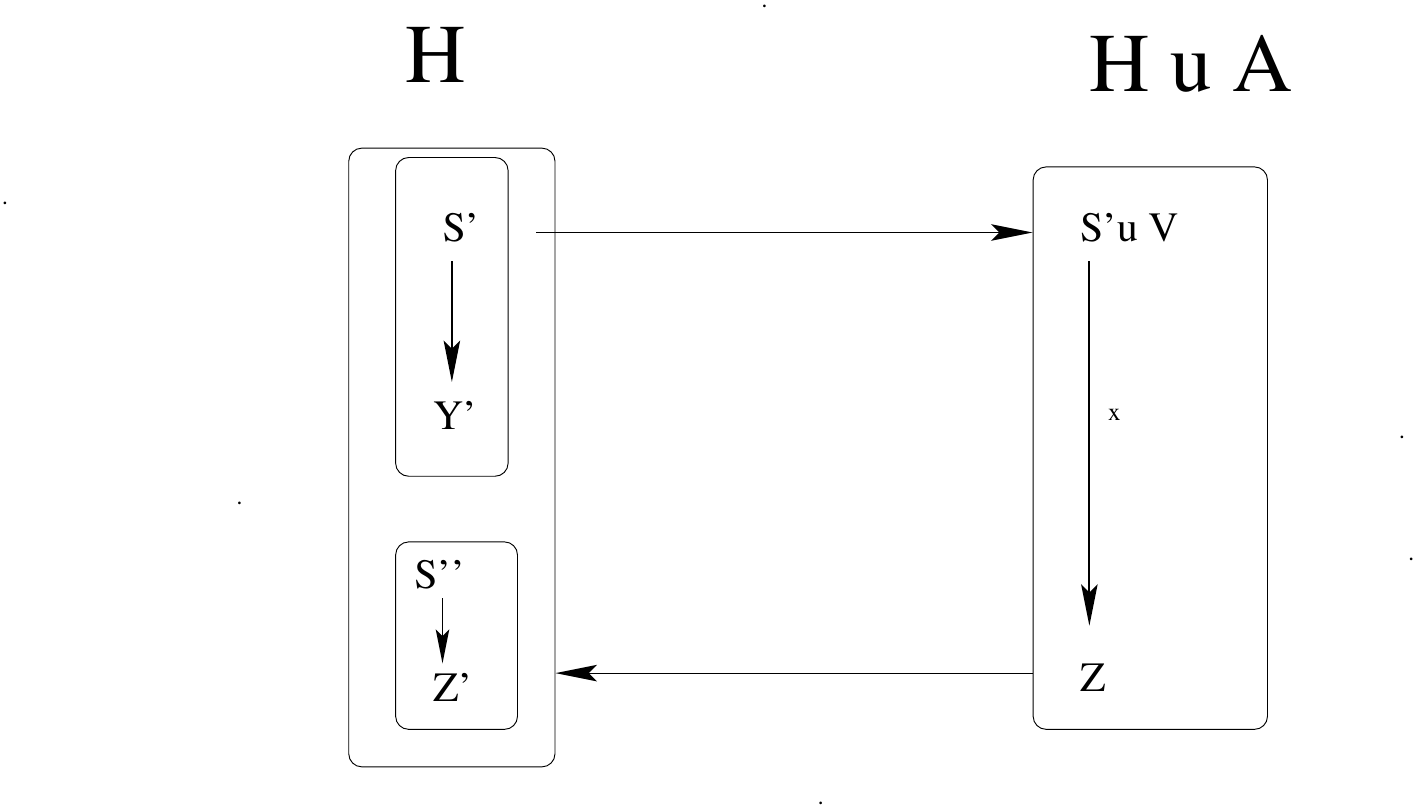}
\caption{ Construction of the function $\kappa$ that maps a non  fast-augmentable sub-solution of the instance $\Pi(H, \gamma)$, denoted $Y'$,   onto   a  fast-augmentable sub-solution  of  $\Pi(H, \gamma)$, denoted $Z'$. The arrow going downward represents the function $\psi$, the arrow from left to right represents an `augmentation', while the arrow from right to left represents the reverse of an `augmentation'.  }
\label{kappa}
\end{figure}   

  So,  let $S'$ be  the feasible set of MISP such that $\psi(S')= Y'$.  First, we have that $S'$ is sub-solution. Indeed, suppose that $S'$ is a solution of MISP, then $\psi$ maps $S'$  to $Y'$ and $Y$. But, since $Y' \not= Y$ as $Y'$ is a sub-solution, we get that $\psi$ is not well-defined. \\
  
   Now, since MISP is fast-augmentable, we have that $S'\cup v$ is a feasible set for some vertex $v$. Without loss of generality, suppose that    $S'\cup v= S$ such that $\psi(S)= X$. Consider a solution $Z$ of $\Pi(G[X], \gamma)$. From the hypothesis, we have that  $Y' \not \subset Z$.  Also, since  $S' \not = S'\cup v$, we have that $Y'\not = Z$. Hence $Z= (Y'- A) \cup B $  where $A \subseteq Y'$ and $B \subseteq (X-Y')$. Finally, since  $Z$ is accessible, there is an element $z \in Z$ such that $Z'= Z-z$ is a feasible set.  That is,  $Z'$  is fast-augmentable since $Z'\cup z= Z$. So define $\kappa(Y')= Z'$.  That is, take $Z'= \psi(S' \cup v) - z$, such that $z \in  \psi(S' \cup v)$ and $\psi(S')= Y'$. See Figure \ref{kappa} for an illustration.\\
   
     The function $\kappa$ runs in polynomial time  since $\phi(H)$ can be found in polynomial time, and  its solution $S'$ can be found and augmented in polynomial time.  Finally $\psi (S'\cup v)$ can be found in polynomial time, and  an element $z$   such that  $\psi(S' \cup v)-z$ is a feasible set can be found in polynomial time.  

\qed\\

However, given a non-fast augmentable solution of  $\Pi(H, \gamma)$  denoted $Y'$,   the existence of a function $\kappa$ that transforms in polynomial time  $Y'$  into a different  solution $Z'$ of $\Pi(H, \gamma)$   entails that the problem $\Pi(G[X], \gamma)$ is slow-accessible. This is proved in the next lemma, which   is  instrumental in showing that there can not be any polynomial function $\kappa$ transforming a non-fast-augmentable feasible set $Y'$ of the Hamiltonian Cycle Problem into a fast-augmentable one. Indeed,  since  Lemma  \ref{HCPnotSlowAccessible} shows that   the  Hamiltonian Cycle Problem  is fast-accessible,  Lemma \ref {kappaEntailsSlowAccessible}  entails that  the Hamiltonian Cycle Problem is not augmentable.\\

\begin{lemma}\label{kappaEntailsSlowAccessible}
Let  $Y' \not \subset Y$, where   $Y'$ is a non-fast-augmentable solution of the sub-instance  $\Pi(H, \gamma)$ of the polynomial time  solvable problem $\Pi(G[X], \gamma)$ and $Y$ is a solution of $\Pi(G[X], \gamma)$. Suppose that  there is a function $\kappa$  that transforms in polynomial time  $Y'$ into a fast-augmentable  sub-solution $Z'$ of $\Pi(H, \gamma)$. Then the problem  $\Pi(G[X], \gamma)$ is slow-accessible.  
\end{lemma}

The proof of Lemma  \ref{kappaEntailsSlowAccessible}  requires the following lemmas \ref{Z'SolutionOfH} and \ref{IfSolutionThenSolution}. 

\begin{lemma}\label{Z'SolutionOfH}

Let  $Y' \not \subset Y$, where   $Y'$ is a non-fast-augmentable solution of the sub-instance  $\Pi(H, \gamma)$ of the polynomial time  solvable problem $\Pi(G[X], \gamma)$ and $Y$ is a solution of $\Pi(G[X], \gamma)$.  Suppose that  there is a function $\kappa$  that transforms in polynomial time $Y'$ into another  feasible set  $Z'$  such that $Z' \not = Y'$ and $Z'$ is fast-augmentable.  Then  $\kappa(Y')$  can be chosen so that  $\kappa(Y')$ and $Y'$  are   both solutions of the sub-instance $\Pi(H, \gamma)$. 
 \end{lemma}
 
 {\bf Proof.}

First,  let $\psi(S')= Y'$ and $\psi(S'\cup v)= Z$ such that $Y'\not \subseteq Z$, and $Y'$ is a solution of $\Pi(H,\gamma)$.  The set  $S' \cup v$ is a feasible set since $S'$ is augmentable.  We aim to show that  there is an independent set of vertices of $\phi(H)$, denoted $S''$, such that   $S'' \not = S'$  and  $\psi(S'')$ is  also a solution of $\Pi(H, \gamma)$,  and  $\psi(S'' \cup w )=Z$ for some element $w \not = v$.\\    

Indeed, let $\phi(H)$ be the graph where $S'$ is a maximal  independent set of vertices. That is, $Y'= \psi(S')$ is a solution  of  $\Pi(H, \gamma)$, where $H$ is a contraction-minor of $G[X]$.   Suppose there is no such a feasible set $S'' \not = S'$.    Since $Z$ is a feasible set, it is accessible. That is, there is an element $z  \in Z$ such that $Z-z$ is a feasible set. Thus,  there must be a maximal independent set of vertices in $\phi(H)$,  $K$, say, such that $\psi(K)= Z-z$, and, by Lemma \ref{claim 0},  $K \subset  S' \cup v$,  since $Z= \psi(S' \cup v)$ and $Z-z \subset Z$. Thus,   either $K= S'$, or  $K \subset S'$ or $K\not = S'$.   If  $K= S'$, then $Y' \subseteq Z$, which is a contradiction. If   $K \subset S'$,  then $K= S' - \{v, v_{_{1}}\}$ for $v_{_{1}} \in S'$.  Thus,   $Z \subseteq Y'$, another contradiction.   Thus $K \not = S' $, and $K$ must  contain the vertex $v$, since it is not a subset of $S'$ but is a subset of $S' \cup v$. \\

This is possible only if   $S'= T \cup W $, where $W=\{ w_{_{1}}, w_{_{2}}, \cdots, w_{_{r}}\}$ is the set of vertices  such that there is a path $v- e_{_{i}}- u_{_{i}} - f_{_{i}}- w_{_{i}}$  in $\phi(G[X])$, where $u_{_{i}}$ is a vertex not in $S'$, $e_{_{i}}$ and $f_{_{i}}$ are edges  in $\phi(G[X])$, and $T= \{ t_{_{1}}, t_{_{2}}, \cdots t_{_{s}}\}$ is the set of vertices of $\phi(G[X])$ where there is no path $v- e_{_{i}}- u_{_{i}} - f_{_{i}}- t_{_{i}}$, as illustrated in Figure \ref{Slow-accessibiltyProof3}.\\
. 

 \begin{figure}[h]
 \center
 \includegraphics[scale=0.8]{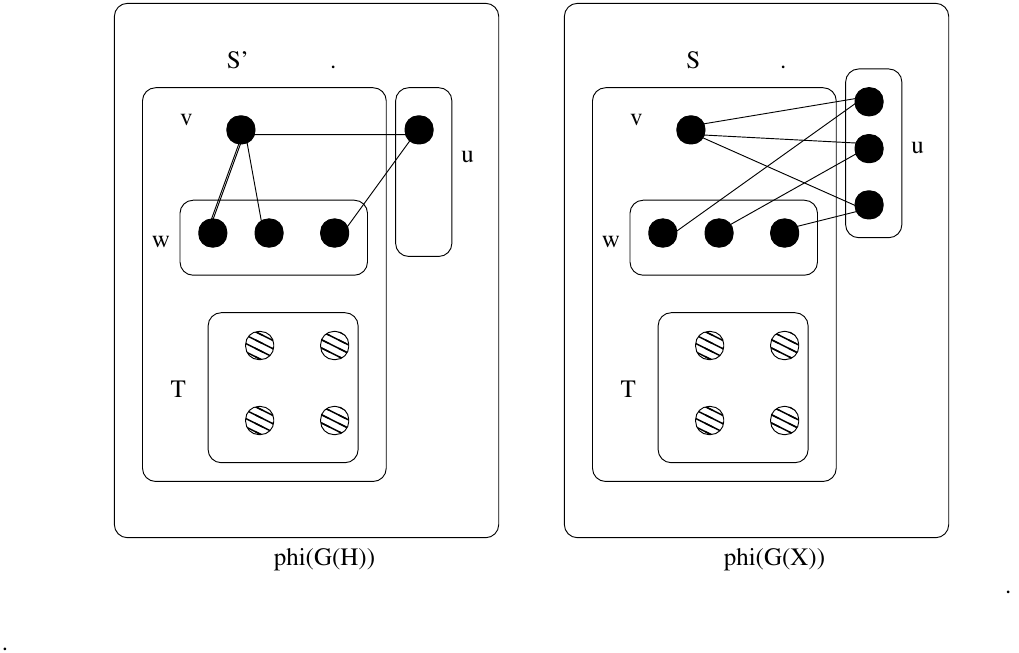}
 \caption{$T= \{t_{_{1}}, t_{_{2}},t_{_{3}},t_{_{4}}\}$ is the set of stripped vertices in the set $S$.  The  feasible set $Y'$ is not fast-augmentable but  the independent set of vertices $S'$ such that $\psi(S')=Y'$ is fast-augmentable as $S'\cup v $.  In $\phi(H)$, $T \cup w_{_{1}} \cup w_{_{2}} \cup w_{_{3}}= S' $ is a maximal set of vertices, while in  $\phi(G[X])$, $T \cup w_{_{1}} \cup w_{_{2}} \cup w_{_{3}} \cup v= S $ is a maximal set of vertices. Note that in $\phi(G[X])$, the edges $e_{_{i}}$ are the edges connecting the vertex $v$ and vertices in $u$. Thus $S'$ is obtained from $S$ by contracting two edges $e_{_{i}}$.   We take $S''= T \cup w_{_{3}} \cup v$.}
 \label{Slow-accessibiltyProof3}
 \end{figure}

 Indeed, since $S'$ is fast-augmentable as $S' \cup v$ and $\psi(S')= Y'$, where $Y'$ is a solution of $\Pi(H, \gamma)$, we have that  the graph $ \phi(H)$ must be such that   $ \phi(H)= \phi(G[X])/ \{e_{_{1}}, e_{_{2}}, \cdots , e_{_{q}}\} $, with $q \leq r$. So that, in the graph  $\phi(G[X]) $,  for all $i \leq r$, the vertices $v$ and and $w_{_{i}}$ are  not adjacent,  and thus are elements of the maximal  independent set $S$. However,     for $i \leq q$, $v$  is adjacent  to  $w_{_{i}}$   in $\phi(H)$ while  we  have  that  in $\phi(H)$ the vertex $v$ and vertices $w_{_{j}}$ for $j> q$ are not adjacent and are thus elements of a maximal independent set. (Notice that $q=2$ in Figure  \ref{Slow-accessibiltyProof3}.)  Hence,   $S''= K=  T \cup \{\ v, w_{_{q+1}}, w_{_{q+2}}, \cdots, w_{_{r}}\}$ is a feasible set in $\phi(H)$, and   $Z'= \psi(S'')$ is a solution of $\Pi(H, \gamma)$.  \\
 
 \qed

\begin{lemma}\label{IfSolutionThenSolution}
Let  $Y' \not \subset Y$, where   $Y'$ is a non-fast-augmentable solution of the sub-instance  $\Pi(H, \gamma)$ of the polynomial time  solvable problem $\Pi(G[X], \gamma)$ and $Y$ is a solution of $\Pi(G[X], \gamma)$. Suppose that  there is a function $\kappa$  that transforms in polynomial time $Y'$ into another solution $Z'$ of $\Pi(H, \gamma)$ such that $Z' \not = Y'$. If $e \in Z'$ and    $Z'-e$ is a solution of $\Pi(H/e, \gamma)$,  then a  proper subset of $Y'$  must also be a solution of $\Pi(H/e, \gamma)$.
\end{lemma}

{\bf Proof.}\\

 From the proof of Lemma \ref{Z'SolutionOfH}, we get $W= \{w_{_{1}}, w_{_{2}}, \cdots, w_{_{r}}\}$,    $S'= T \cup \{w_{_{1}}, w_{_{2}}, \cdots, w_{_{r}}\}$ and  $S''=  T \cup \{\ v, w_{_{q+1}}, w_{_{q+2}}, \cdots, w_{_{r}}\}$, where $w_{_{i}}$ is not adjacent to $v$ in $\phi(H)$ for $i>q$.  See an illustration in Figure  \ref{Slow-accessibiltyProof3}, where $q=2$ and $r=3$.  Now, if $Z'-e$ is a solution of $\Pi(H/e, \gamma)$,  then, by Lemma \ref{claim 0} ,  either $Z'-e= \psi(S''-t)$ for some $t \in T$, or  $Z'-e= \psi(S''-w_{_{i}})$ for $i$ such that $ q+1 \leq i \leq r$, or  $Z'-e= \psi (S''-v)$. (In Claim \ref{OnlyOneElement} we claim that only one element of $S''$ should be removed, lest $\phi$ is not well-defined. However, it is worth noting that the argument is not altered if we accept that more than one element of $S''$ is removed concomitantly with the contraction by $e$.)  \\ 
 
 a) First we note that  $T \cup \{v\} \cup W$ is  not an independent set of vertices in graph  $ \phi(H)$ since  some vertices in  $W$  are adjacent to $v$ in the graph $ \phi(H)$.    Suppose that   $Z'-e= \psi(S''-v)$.  Since  only the vertex $v$ is removed  from the    set $S''$   when the edge $e$ is contracted from $H$, we have that  $T\cup W  $ is still a maximal  independent set of vertices  in $\phi(H/e)$,  and since  $T \cup W  = S'$,  we get that  $Y'$  is a solution  of $\Pi(H/e, \gamma)$.   However, we also get that   $Z'- e = \psi (T \cup W ) $, since $T \cup \{ w_{_{q+1}}, w_{_{q+2}}, \cdots, w_{_{r}} \}$ is not a maximal independent set of vertices in $\phi(H/e)$ but  $T \cup W$ is a maximal independent set of vertices in $\phi(H/e)$. Hence, either    $Z' - e = Y'$, or $\psi$ is not well-defined as it would match  $T \cup W$ to two different sets $Z'-e$ and $Y'$.  Both are contradictions. Therefore, if $Z'-e$ is a solution of  $\Pi(H/e, \gamma)$, then either $t$  or $w_{_{i}}$  for $i$ such that $ q+1 \leq i \leq r$ is removed from $S''$.\\
 
b)  Suppose that  $Z'-e= \psi(S''-t)$ or $Z'-e= \psi(S''-w_{_{i}})$,  for $i$ such that $ q+1 \leq i \leq r$. We then get the sets $(T\cup W)-t$ or $(T\cup W)-w_{_{i}}$ which are both maximal independent sets of vertices  in $\phi(H/e)$, since $v$ is adjacent to vertices $w_{_{j}}$ such that $1 \leq j\leq q$.  Since  both $(T\cup W)-t$ and  $(T\cup W)-w_{_{i}}$ are subsets of  $ S'= T\cup W$, we  get that  a subset of $Y'$ is a solution of $\Pi(H/e,\gamma)$, by Lemma \ref{claim 0}. \\

\qed

\begin{claim} \label{OnlyOneElement} 
Suppose that  $e \in Z'$ and    $Z'-e$ is a solution of $\Pi(H/e, \gamma)$,  then  $Z'-e= \psi(S''-A)$, where $A$ must be a singleton.
\end{claim}

{\bf Proof.}
Suppose that $A$ contains more than one element. So, without loss of generality, suppose that  $Z'-e= \psi(S''-\{s_{_{1}}, s_{_{2}} \})$. Since $S''$ is a maximal independent set of vertices in $\phi(H)$, we have that  the vertices $s_{_{1}}$ and $ s_{_{2}}$ can not be adjacent  in  the graph   $\phi(H)$. That is, in the worst case,  there must be  a path   $s_{_{1}}-e_{_{1}}-u-e_{_{2}}-s_{_{2}}$, where the vertex  $u \not \in S''$. Hence to remove both the vertex $s_{_{1}}$ and $s_{_{2}}$ necessitates  to contract at least the edges  $e_{_{1}}$ and $e_{_{2}}$.\\

 Now,  suppose that   $s_{_{1}}= \phi(a)$ and $s_{_{2}}=\phi(b)$ for some elements $a$ and $b$ in the graph $H$. Since $H$ is contracted by a single element $e$, we may suppose that the elements $a$  and $s_{_{1}}$  are  removed from the feasible sets $Z'$ and $S''$, respectively. If $s_{_{2}}$  is also removed from $S''$, it means that $\phi(b)$ does not exist. Thus, $\phi$ is not well-defined. 
\qed

\vspace{3mm}

{\bf Proof of Lemma \ref{kappaEntailsSlowAccessible}.}

We first recall that, given the input $G[X]$ of a polynomial time solvable problem $\Pi(G[X], \gamma)$, the graph $\phi(G[X])$  is an instance of the search problem  MISP. And, a solution of $\phi(G[X])$ is a maximal independent set  of vertices in the graph $\phi(G[X])$.\\

To prove Lemma \ref{kappaEntailsSlowAccessible},  we have to show that, if the conditions of the hypothesis are satisfied, then, for  every solution $Y$ of any   instance $G[X]$,  every  accessibility chain  $ \emptyset \unlhd  I_{_{1}} \unlhd I_{_{2}} \unlhd \cdots, I_{_{i}} \unlhd \cdots \unlhd Y$  must contain   a move from the feasible set  $I^{(i)}$ to $I^{(i-1)}$, for  some $i$, where moving  from the feasible set  $I^{(i)}$ to the feasible set  $I^{(i-1)}$  requires two operations of contraction.\\
  
Indeed, let $\kappa(Y')= Z'= \psi(S' \cup v) - z$, where  $z \in Z=  \psi(S' \cup v)$, and $\psi(S')= Y'$, as shown in part (B) of the proof of Theorem \ref{MainTheorem}. Let $S=S' \cup v= (T \cup  W) \cup v$,  $S''= T \cup \{w_{_{q+1}}, w_{_{q+2}}, \cdots , w_{_{r}}\}\cup v$ ,  where  $W=\{ w_{_{1}}, w_{_{2}}, \cdots, w_{_{r}}\}$ is the set of vertices of $\phi(G[X])$  such that there is a path $v- e_{_{i}}- u_{_{i}} - f_{_{i}}- w_{_{i}}$  in $\phi(G[X])$, where $u_{_{i}}$ is a vertex not in $S'$, $e_{_{i}}$ and $f_{_{i}}$ are edges  in $\phi(G[X])$, and $T= \{ t_{_{1}}, t_{_{2}}, \cdots t_{_{s}}\}$ is the set of vertices of $\phi(G[X])$ where there is no path $v- e_{_{j}}- u_{_{j}} - f_{_{j}}- t_{_{j}}$, as illustrated in Figure \ref{Slow-accessibiltyProof3}. For an illustration of the unfolding proof, see Figure \ref{Slow-accessibiltyProof2} , where  we may assume that $W$ contains a unique element $w$  so that  $Y'= \psi(S')$  where $S'= T \cup w$, and  $Z'= \psi(S'')$, where   $S''= T \cup v$.   \\

 \begin{figure}[H]
 \center
 \includegraphics[scale=0.55]{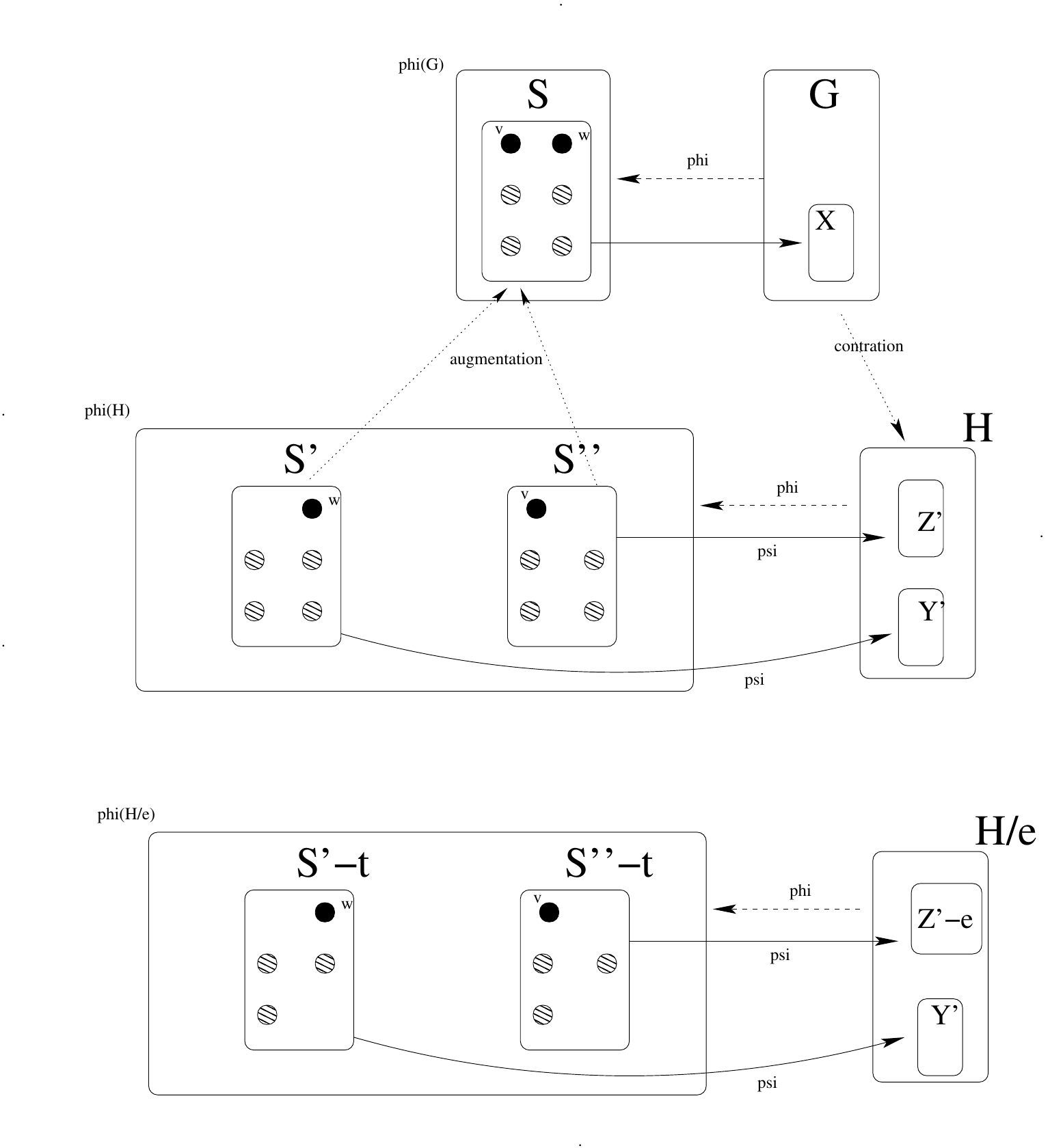}
 \caption{$T= \{t_{_{1}}, t_{_{2}},t_{_{3}},t_{_{4}}\}$ is the set of stripped vertices in the set $S$.  The  feasible set $Y'$ is not fast-augmentable but  the independent set of vertices $S'$ such that $\psi(S')=Y'$ is fast-augmentable as $S'\cup v $.  Notice that $Y'= \psi(T \cup w)$ and $Z'= \psi (T \cup v)$.   }
 \label{Slow-accessibiltyProof2}
 \end{figure}

Suppose  $X$ is a set of edges. Then,  since   $Z'=\kappa(Y')$  is a feasible set in the graph $H$,  there is an edge $e$ in  $Z'$  such that
 $Z'-e$ is a feasible set (although $Z'-e$  may not necessarily be a solution of the problem  $\Pi(H/e, \gamma)$). 
  If $X$ is a set of vertices, then, since   $Z'=\kappa(Y')$  is a feasible set in the graph $H$, there is a vertex  $a$ in  $Z'$  such that
 $Z'-a$ is a feasible set. Thus, the vertex $a$ is an end-vertex of an edge $e$ such that contracting the graph $H$  by the edge $e$ only removes the vertex $a$ from the feasible set $Z'$.  So, either $X$ is a set of edges or edges of vertices,  consider  the graph $H/e$. \\
 
   {\bf Nota}: In what follows the proof only considers the case where $X$ is a set of edges. However,  it suffices   to change $Z'-e$ to $Z'-a$ to fit the proof for the case where $X$  is a set of vertices. Notice that in both cases,  the notation $H/e$ carries   the same  concept.  \\
 
 Suppose that $Z'-e$ is a solution of  $\Pi(H/e, \gamma)$.  Then,  by Lemma \ref{IfSolutionThenSolution}, we have that a proper subset of $Y'$ must also be a solution of   $\Pi(H/e, \gamma)$. So,  suppose there is a set $A$ of  elements of $Y'$ such that $Y'-A$  is a solution  of $\Pi(H/e, \gamma)$.   Since $Y'-A$ is a solution of $\Pi(H/e, \gamma)$,  and $A-e$ is a set of  edges of  $H/e$, we have that   $A-e \subset cl(Y'-(A-e))$ in $H/e$.   That is,   contracting by the edge $e$ automatically remove $A$ from the feasible set  $Y'$.  This is possible only if  $Z'-e= \psi(S''-t)  $ for some element $t \in T$ or  $Z'-e= \psi(S''-w_{_{i}})  $ for  $ q+1 \leq i \leq r$ (many such vertices may be removed at the same time).  Indeed,  suppose that  contracting by $e$ removes the element $v$ from $S''$  instead of  vertices $t$ or $w_{_{i}}$. Then $S''-v= T\cup \{ w_{_{q+1}}, w_{_{q+2}}, \cdots , w_{_{r}}\}$  can not be a maximal independent set of vertices in $\phi(H/e)$  since    $T\cup W= S'$   is a maximal independent set of vertices in $\phi(H/e)$.  Thus, by Lemma \ref{claim 0},   $Z'-e$ is not a feasible set in $\Pi(H/e, \gamma)$.  This is  a contradiction.   \\  
 
  {\bf Nota: } If $X$ is a set of vertices, we say: Since $Y'-A$ is a solution of $H/e$,  and $A$ is a set of  vertices of  $H/e$, we have that   $A \subset cl(Z'-A)$ in $H/e$.\\

The remainder of the proof, which mimics Example \ref{Matching2}, consists of showing that, given $Y'$ and $Z'$ such that $Z'= \kappa(Y')$ and both $Y'$ and $Z'$ are solutions of $\Pi(H, \gamma)$, if one contracts sequentially the edges  $\{ e_{_{1}}, e_{_{2}}, \cdots e_{_{s}}\}$ of $H$, there  is an ordering of these edges $e_{_{i}}$, for $1\leq i \leq s$, such that every contraction up to some $e_{_{\omega}}$ yields  a sequence of  feasible sets  that are  proper subsets  of  $Z'$ and $Y'$, respectively. That is,  each  subset must be different     from one  contraction to the next  one. However, there will be an edge $e_{_{\omega+1}}$ such that the subset from $Z'$ is different from the preceding one, while  the subset from $Y'$ is the same as the preceding one. Thus, by Lemma \ref{IfSolutionThenSolution}, the subset from $Z'$ cannot be a feasible set.   Therefore, one would need at least  two consecutive contractions to change the subset from $Z'$ into a feasible set.      \\
 
 Indeed,  suppose  $X$ is a set of  edges.  Since  $Z'$ is accessible,  there is  an accessible chain   from $\emptyset$  to $ Z'$. That is, there is a sequence of edges $E_{_{i}}= (e_{_{1}},  e_{_{2}}, \cdots, e_{_{i}})$, where  $i \leq n$ and  $n=|Z'|$, such that $Z'-E_{_{i}}$ is a feasible set for all $i \leq n$.   This entails  that  there is a subsequence $E_{_{\rho}}$,  with $\rho \leq n$,  such that   $E_{_{\rho}}= (e_{_{1}},  e_{_{2}}, \cdots, e_{_{\rho}})$  is  a sequence of   edges $e_{_{i}} \in Z'$  and $Z'-E_{_{i}}$, for $1 \leq i \leq \rho$, is a feasible set  by being   a solution of $\Pi(H'/ \{ e_{_{1}}, \cdots , e_{_{i}}\})$. ($Z'-E_{_{i}}$ may be a feasible but not a solution of $\Pi(H'/ \{ e_{_{1}}, \cdots , e_{_{i}}\})$). By  Lemma \ref{IfSolutionThenSolution},    $E_{_{\rho}}= (e_{_{1}},  e_{_{2}}, \cdots, e_{_{\rho}})$ is also  a sequence of   edges  such that  there is a subset $A_{_{i}} \subset Y'$, $A_{_{i}} \not = \emptyset$,  and  $ Y'-A_{_{i}}$ is also a feasible set  by being  a solution of $\Pi(H'/ \{ e_{_{1}}, \cdots , e_{_{i}}\})$.  \\
  
  Similarly, if $X$ is a set of vertices,  let $V_{_{\rho}}= (a_{_{1}},  a_{_{2}}, \cdots, a_{_{\rho}})$ be a sequence of  the vertices $a_{_{i}} \in Z'$ such that, for $1 \leq i\leq \rho$, $Z'-V_{_{i}}$ is a feasible set  by being a solution of $\Pi(H'/ \{ e_{_{1}}, \cdots , e_{_{i}}\})$.    Let  $E_{_{\rho}}= (e_{_{1}},  e_{_{2}}, \cdots, e_{_{\rho}})$ be a sequence of  the edges in $H$ such that contracting $H$ by $e_{_{i}}$ removes the  vertex $a_{_{i}}$ from the feasible set $Z'_{_{i}}$. By Lemma   \ref{IfSolutionThenSolution},    $E_{_{\rho}}= (e_{_{1}},  e_{_{2}}, \cdots, e_{_{\rho}})$ is also the sequence of  edges in   the graph $H$   such that  there is a subset $A_{_{i}} \subset Y'$, $A_{_{i}} \not = \emptyset$, such that  $ Y'-A_{_{i}}$ is also feasible set  by being a solution of $\Pi(H'/ \{ e_{_{1}}, \cdots , e_{_{i}}\})$. \\
   
   Let $Z'_{_{i}}=   Z'- \{{e_{_{1}}, e_{_{2}}, \cdots, e_{_{i}}}\}$ if $X$ is a set of edges, (or let  $Z'_{_{i}}=   Z'- \{{a_{_{1}}, a_{_{2}}, \cdots, a_{_{i}}}\}$ if $X$  is a set of vertices). Let $Y'_{_{i}}= Y'-A_{_{i}}$,  let $K_{_{i}}= H'/{\{{e_{_{1}}, e_{_{2}}, \cdots, e_{_{i}}}\}}$, let $G_{_{i}}= \phi(K_{_{i}})$,  let $S'_{_{i}}$ and $S''_{_{i}}$ be maximal independent sets of vertices in $G_{_{i}}$. That is,   $S'_{_{i}}$ is  the maximal  independent set of vertices such that $Y'_{_{i}}= \psi(S'_{_{i}})$, and   $S''_{_{i}}$ is  the maximal  independent set of vertices such that $Z'_{_{i}}= \psi(S''_{_{i}})$. Let $T_{_{i}}$, $W_{_{i}}$ and $W'_{_{i}}$    be the  vertices from $T$,  $W$ and $\{w_{_{q+1}}, w_{_{q+2}}, \cdots, w_{_{r}}\}$, respectively,  that are left over   after contracting  the edge $e_{_{i}}$ from the graph $H$ in the sequence $E_{_{i}}$. \\

    Consider the feasible set $Z'_{\rho}$. Then $Y'_{\rho}$ is also a feasible set by Lemma \ref{IfSolutionThenSolution} (Recall the  definition of the sequence  $E_{_{\rho}}$).  So, we get that $Z'_{\rho}= \psi (v \cup T_{\rho} \cup W_{\rho}')$.  Indeed, for  $i \leq \rho$,  as mentioned above, contracting by $e_{_{i}} $ would never remove the vertex  $v$ since $T_{i} \cup W_{i}$ would be a maximal independent set of vertices in $\phi(K_{_{i}})$, whence  $Z'_{_{i}}$ would not be a feasible set. This is a contradiction. \\

     Finally suppose that $e_{\rho+1}$ is the first element in the sequence $E_{_{n}}$  such that $ Z'- \{{e_{_{1}}, e_{_{2}}, \cdots,  e_{_{\rho}},e_{_{\rho+1}}}\}= \psi(T_{\rho +1} \cup W'_{\rho+1})$. That is, $e_{\rho+1}$ is the  first element in the sequence of contractions whose removal removes the vertex $v$ from $S''_{\rho}$.\\
     
       Such an element $e_{_{\rho+1}}$ must exist. Indeed, suppose that $e_{_{\rho+1}}$ does not exist.  Then  all the elements $e_{_{i}} \in E_{_{n}}$ can be removed from $Z'$ until one gets the empty set. However,    after removing  from $Z'$ all the elements $e_{_{i}}$, $1 \leq i \leq n$ , we are left with the maximal independent sets of vertices  $S'_{_{n}}= \{ \text{some vertices $t$ and  $w$}\}$ or  $S''_{_{n}}= \{ v, \text{some vertices $t$ or $w$}\}$.   Now, since  $Z'_{_{n}}= \emptyset$  then $\phi$ is not well-defined, as it matches   $K_{_{n}}$ to both the Empty Graph $G[\emptyset]$   and the graph $G_{_{n}}$ such that $S''_{_{n}}$ is a maximal set of independent vertices in $G_{_{n}}$. See Figure  \ref{Slow-accessibiltyEMPTY} for an illustration. \\
       
        Indeed, by Definition 1.1, we have that  $\phi$ matches the Empty Graph to the Empty Graph  and $\psi$ matches the Empty Set to the Empty Set.  So,  by Definition 1.1, we have that $\emptyset= \psi(T_{n} \cup W'_{n} \cup v)$ if and only if  $T_{n} \cup W'_{n} \cup v$ is a solution of  $G_{_{n}}= \phi(K_{_{n}})$, and the empty set  $\emptyset$ is a solution of $K_{_{n}}$.  Now, since  the empty set $\emptyset$  is a solution of the empty graph $G[\emptyset]$,    we also  have  that  $\emptyset= \psi(\emptyset )$ if and only if  $\emptyset $ is a solution of $\phi(G[\emptyset])$, where $G[\emptyset]$ is the Empty Graph. Hence $\phi$ is not well-defined as it matches $G[\emptyset]$ to two different instances of the MIS Problem. Indeed  $G_{_{n}}$ is not the Empty Graph since  it contains the independent set of vertices $T_{n} \cup W'_{n} \cup v$.  \\
       
  \begin{figure}[h]
 \center
 \includegraphics[scale=0.6]{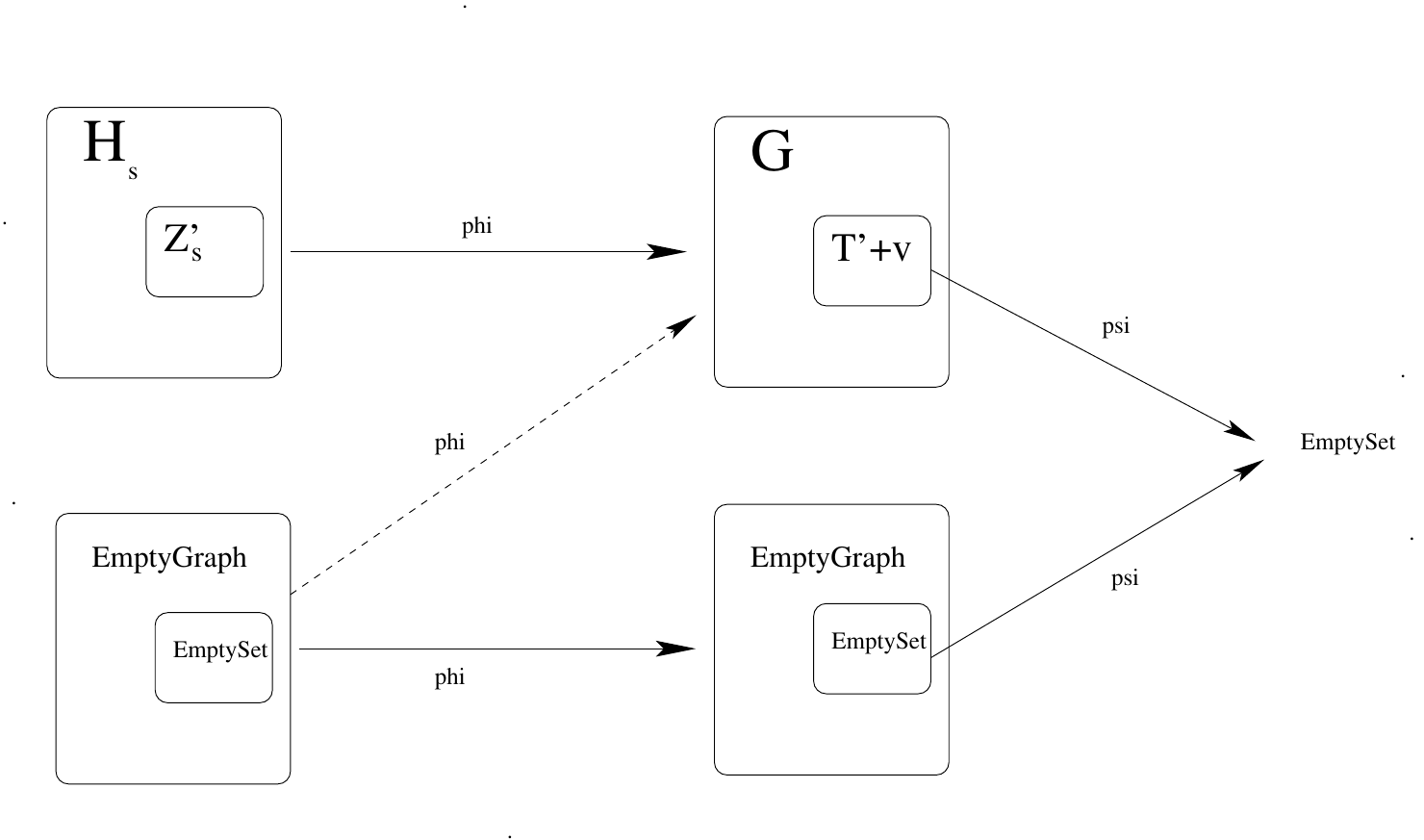}
 \caption{We have that $\emptyset= \psi(T_{_{n}} \cup W'_{_{n}} \cup v)$ if and only if  $T_{_{n}} \cup W'_{_{n}} \cup v$ is a solution of  $G_{_{n}}= \phi(G[\emptyset])$.  We also have that  $\emptyset= \psi(\emptyset )$ if and only if  $\emptyset $ is a solution of $\phi(G[\emptyset])$, where $G[\emptyset]$ is the Empty Graph. Hence $\phi$ is not well-defined as it matches $G[\emptyset]$ to two different instances of the MIS Problem. }
 \label{Slow-accessibiltyEMPTY}
 \end{figure}

      Now,  given the element  $e_{_{\rho+1}}$,  we  get  that  the graph $\phi(K_{_{\rho+1}})$  contains the maximal independent set    $ S'_{_{\rho+1}}= T_{_{\rho+1}} \cup W_{\rho+1}= S'_{_{\rho}} $, since the vertex  $v$ is the only vertex  removed from $S''_{_{\rho}}$ . Hence,    $Y'- A_{_{\rho}}$  is also a solution of  the problem $\Pi(K_{_{\rho+1}}, \gamma)$ since $Y'- A_{_{\rho}}= \psi(S'_{_{\rho+1}})$. This contradicts  Lemma \ref{IfSolutionThenSolution}, which claims that  a solution of  $\Pi(K_{_{\rho+1}}, \gamma)$  should  rather be  $Y'- A_{_{\rho+1}}$.  Thus,   $ Z'_{_{\rho+1}}$  can not be a  solution of the problem $\Pi(K_{_{\rho+1}}, \gamma)$.  (Indeed, $S''_{\rho +1}= S''_{\rho }- v $ is not a maximal set of vertices since it is then contained in $S'_{\rho +1}.$ )  \\

  However,  since  $Z'_{_{\rho+1}}$ is a feasible set, it must be a solution of a contraction-minor of $H$ other than $K_{_{\rho+1}}$. This is possible only  if   $Y'-A_{_{\rho}}$ is not a solution of that same contraction-minor, by Lemma \ref{IfSolutionThenSolution}.  This is possible   only if   vertices $w_{_{i}}$, for $1 \leq i \leq q$, are also  removed from $S'_{_{\rho+1}}$. Without loss of generality, we may suppose that the vertices  $w_{_{i}}$, for $1 \leq i \leq q$ are removed in the order $1,2,\cdots, q-1$. (We recall that $S'= T \cup \{ w_{_{1}}. \cdots w_{_{r}}\}$  and $S''= T \cup \{ w_{_{q}}. \cdots w_{_{r}}\}\cup v$, with $q >1$. Thus,  after removing $v$ from  $S''_{_{\rho}}$, we get that   $S'_{_{\rho}} =  S'_{_{\rho+1}}$, and it  a maximal set of vertices in $G_{_{\rho+1}}$.)    However, by Lemma \ref{claim 0}, removing  the  vertex $w_{_{1}}$ from $S'_{_{\rho +1}}$ requires to contract $K_{_{\rho+1}}$ by  at least one other edge. Indeed,    $S'_{_{\rho+1}}-w_{_{1}}= S'_{_{\rho+2}}$ must be the solution of a contraction-minor of the graph $G_{_{\rho+2}}$  such that  $\phi(K_{_{\rho+2}})=  G_{_{\rho+2}}$. Thus, $K_{_{\rho +2}}$ must also be a contraction-minor of $K_{_{\rho+1}}$. \\ 
  
  Finally, since $Z'_{_{\rho+1}}$ can be a  maximal independent  set of vertices only after removing  all the  vertices $w_{_{i}} \in \{ w_{_{1}}, \cdots, w_{_{q}}\}$, and for each such vertex $w_{_{i}}$ at least  one extra edge $f_{_{j}}$  must be contracted, we get that  there must be some edge $f_{_{1}}, \cdots, f_{_{q}} $, with $f_{_{j}} \not = e_{_{i}}$ for $ j \leq q,   i \leq \rho+1$, such that  $Z'_{_{\rho+1}}$  is a solution of the problem $\Pi(K_{_{\rho}}/\{\ e_{_{\rho+1}}, f_{_{1}}, \cdots, f_{_{q}}\}, \gamma)$. That is, the transition from the feasible set $Z'_{_{\rho}}$ to $Z'_{_{\rho+1}}$ requires more than one  contraction.  Hence the problem $\Pi(G[X], \gamma)$ is slow-accessible.\\

\vspace{5mm}

\qed\\

\vspace{5mm}


\item{\bf Sufficiency.}
 Suppose now that  Augmentability holds. Define an algorithm that solves the problem $\Pi(G[X], \gamma)$  in polynomial time  as follows. The algorithm consists of building a solution by moving from a feasible set  to another by augmentation. \vspace{3mm}




{\bf Algorithm $\mathcal{B}$: Generalised Greedy Algorithm}\label{Generalised Greedy Algorithm}

Consider a problem $\Pi$, where the input $X$, the vertex-set or edge-set of an isthmus-less  connected  labelled graph $G$, contains $n$ elements. Let $\kappa$ be a function from $\mathcal{I}$ to $\mathcal{I}$ that runs in polynomial time.  The function  $\kappa$ can be the Identity function. 
\begin{itemize}
\item Step 1. Let $i=0$ and let $Y^{(0)}= \emptyset$. (We can do that since $\emptyset$ is a feasible set.)
\item Step 2. Amongst all elements of $X - \kappa(Y^{(i)})$, choose an element $x$ such that  $\kappa(Y^{(i)}) \cup x$ is a sub-solution of the problem. By Axiom M2', such an element $x$ exists if $Y^{^{(i)}}$ is a sub-solution   (not basis) of $\Pi(G[X])$. If no such an element $x$ exits, stop, output $Y= \kappa(Y^{^{(i)}})$. 
\item Step 3. Let $Y^{^{(i+1)}}=  \kappa(Y^{^{(i)}}) \cup x$. Go to Step 2. 
\end{itemize}  

Algorithm $\mathcal{B}$ must eventually  terminate and outputs a solution of the problem $\Pi$. Indeed, since  $X$ contains a finite number of elements, Step 2 would eventually exhaust all the elements $x$ such that $Y^{(i)} \cup x$ is a sub-solution.  Moreover, since by Accessibility there are paths from $\emptyset$ to the    solutions, a solution would eventually be reached.\\

Finally Algorithm $\mathcal{B}$ runs in polynomial time.  Indeed, since $Y^{^{(i)}}$ is an augmentable  feasible set, there is a contraction-minor $G^{^{(i)}}$ of $G$,  such that $Y^{^{(i)}}$ is a solution of $\Pi$ instanced on $G^{^{(i)}}$, and  there is an element $x$  such that $Y^{^{(i)}} \cup x$ is a feasible set, there is a graph, $G^{^{(i+1)}}$ such that $G^{^{(i)}}$  is a contraction-minor of $G^{^{(i+1)}}$, and $Y^{^{(i)}} \cup x$ is a solution of $\Pi$ instanced on $G^{^{(i+1)}}$.\\

 Now, to construct $G^{^{(i+1)}}$ from $G^{^{(i)}}$, it suffices to check amongst  $|X|-|Y^{^{(i)}}|$  elements which one, if added  by re-insertion to $G^{^{(i)}}$, yields a graph whose one solution is $Y^{^{(i)}} \cup x$.  And checking whether $Y^{^{(i)}} \cup x$ satisfies property  $\gamma$ in $G^{^{(i+1)}}$ consists of checking the incidence properties of vertices and edges of $Y^{^{(i)}} \cup x$ and $G^{^{(i+1)}}$. But, since we  already know   the incidence properties of vertices and edges of $Y^{^{(i)}}$ and $G^{^{(i)}}$, checking  the incidence properties of vertices and edges of $Y^{^{(i)}} \cup x$ and $G^{^{(i+1)}}$  consists only of checking how the extra element $x$ (which may be an edge  or a vertex) modifies the incidence properties of vertices and edges of $Y^{^{(i)}}$ and $G^{^{(i)}}$. Suppose that $G^{(i)}$  contains $m$ elements, with $m \leq n$. The worst case would occur if the checking  takes a time exponential in $m$ and $1$ (the new element added). Thus the worst case would be $\mathcal{O}({m^1})$ or $\mathcal{O}({1^m})$. Hence this can be done in polynomial time. That is, each iteration adding an element $x$ can be performed in  time polynomial in $n$.\\

 Finally,  suppose that a solution contains  at most $k$ elements. Then the algorithm $\mathcal{B}$ would  run in  at most $k$ iterations, where every iteration takes a time that is polynomial in $n$.  

\end{enumerate}
\qed\\

\begin{lemma}\label{HCPnotAugmentable3}
The Hamiltonian Cycle Problem is not augmentable.
\end{lemma}

{\bf Proof.}
 By Lemma \ref{CycleNotAugmentable},  if $C'$ is a  non-Hamiltonian cycle in $G$, then $C'$ is a feasible set of HCP that is not fast-augmentable,   and there is no Hamiltonian cycle  $C$ such that $C' \subset C$. By Lemma \ref{HCPnotSlowAccessible}, the HCP problem is not slow-accessible. Thus, by Lemma  \ref{kappaEntailsSlowAccessible}, there can be no polynomial time computable function  $\kappa$ that transforms a non-fast augmentable sub-solution of HCP into a fast-augmentable one. Hence, HCP does not satisfy Axiom M2'.   

\qed\\

\begin{corr}\label{pnotnp}
\[\mathcal{P}\not = \mathcal{NP} \]
\end{corr}

{\bf Proof}

By Lemma \ref{HamiltonianAccessibility}, the Hamiltonian Cycle Problem is accessible, but by Lemma  \ref{HCPnotAugmentable3}, the Hamiltonian Cycle Problem does not satisfy Axiom M2'.  Hence,  it is not solvable in polynomial time.  Therefore, by Theorem \ref{HCeqHatHC}, the decision problem  $\hat{HC}$ is not  in $\mathcal{P}$. Hence  $\mathcal{P}\not = \mathcal{NP} $.

\qed\\

\pagebreak

\section {APPENDICES} 

\begin{example} \label{exampleContraction} {\bf Contraction and deletion of edges}
\end{example}

Let $G(V,E)$  be a  connected labelled  graph with vertex-set $V$ and edge-set $E$.  Labelling $V$ and $E$ means that every edge and vertex can be indexed so that  $V= \{ v_{_{1}}, \cdots, v_{_{n}}\}$ and   $E= \{ e_{_{1}}, \cdots, e_{_{m}}\}$.   The \text{deletion} of the edge $e \in E$ consists of removing the edge $e$, and leave everything else unchanged, to obtain the graph  $H(E',V')$, where  $V' =V$, and $E'= E-e$.  We denote $H$ as $G\setminus e$.   A \textit{deletion-minor} of $G$  is the graph $G \setminus A $, where  $A \subseteq E$. Note that the order of deletions is irrelevant. That is, for any two edges $e$ and $f$, $G \setminus e \setminus f = G\setminus f \setminus e$.\\

\begin{figure}[h]
\center
\includegraphics[scale=0.35]{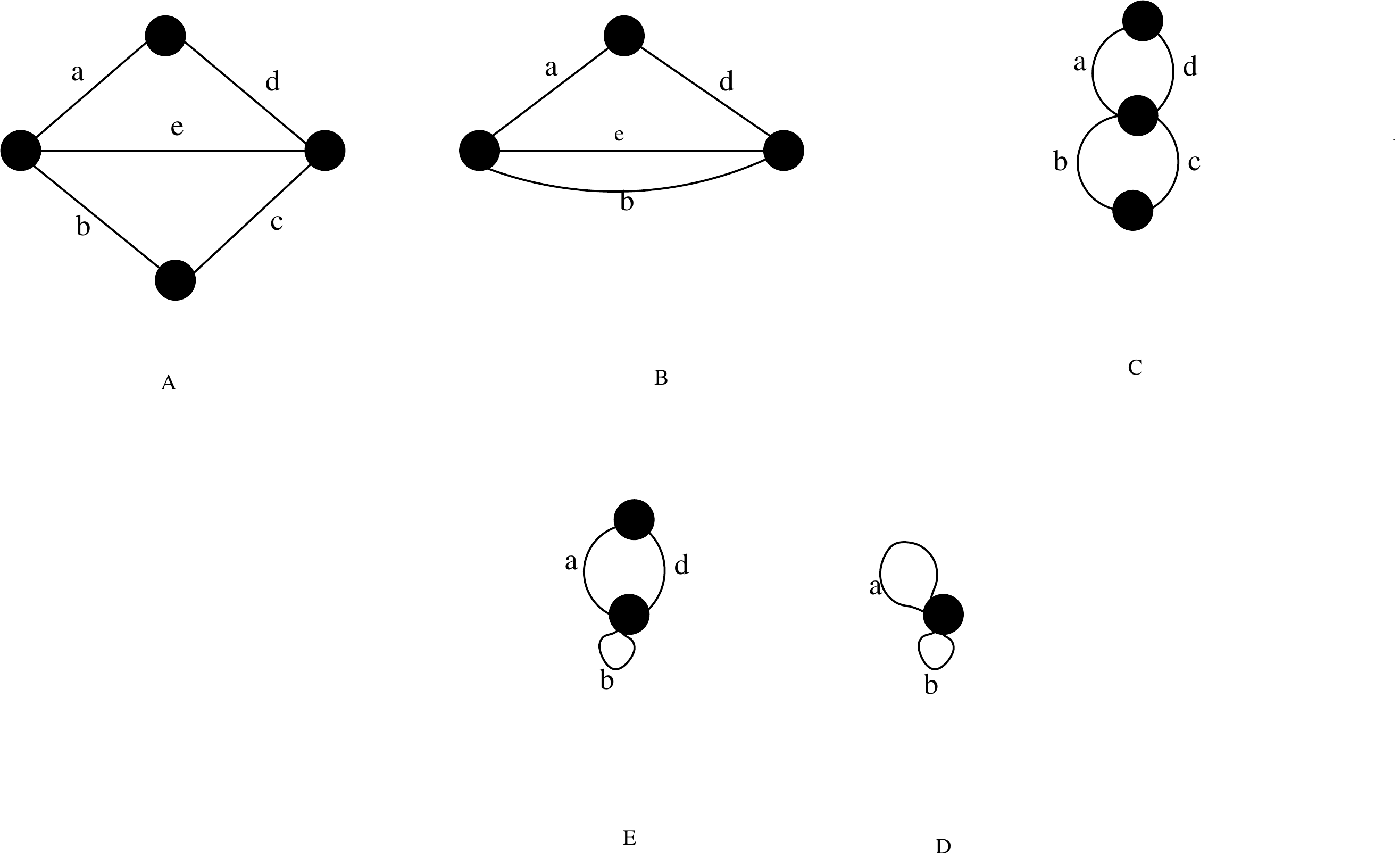}
\caption{A: graph $G$; B: $G/ \{c\}$; C: $G /\{e\}$ ; D: $G/\{c,d,e\}$; E: $G/\{e, c\}$;  Going from the graph  A to the graph  B is a contraction by the edge $c$, while going from B to A is the re-insertion  of the edge $c$.  }
\label{minors}
\end{figure}

  Let $e \in E$, where $e=\{v_{_{i}}, v_{_{j}}\}$ and $i \not = j$. The \textit{contraction}  of the edge  $e$ consists of  deleting the edge $e$, and merging its end-vertices $v_{_{i}}$ and  $v_{_{j}}$ into a single vertex. The new vertex formed by merging $v_{_{i}}$  and $v_{_{j}}$ takes one  of the merged labels.    If $e$ is a loop ($i=j$), then contracting  $e$ consists of deleting $e$.  If $G$ consists of a single vertex and   a loop $e$, contracting   $e$ consists of deleting the single vertex and  the loop. One thus   obtains  the \textit{empty graph}, that is, the  graph with $E=\emptyset$ and $V= \emptyset$. We denote the new graph  obtained from $G$ after contracting  the edge $e$  as $G/e$.  If $e$ is not a loop and  $H(E',V')= G/e$, then $V' = V-v$, and $E'=E-e$, where $v$ is one of the end-vertices of $e$.  Let $A \subseteq E$. The graph $H$ is a \textit{contraction-minor} of $G$ if $H= G /A $. That is, $H$ is obtained from $G$   by  contracting the  edges in $A$. Note that the order of contractions is irrelevant. That is, for any two edges $e$ and $f$, $G/e/f = G/f/e$.  See Figure \ref{minors}   for an  illustration. \\

 Note also that  the order of deletion and contraction is irrelevant.  That is, for any two edges $e$ and $f$, $G \setminus e / f = G / f \setminus e$.  A graph $H(E',V')$  is a \textit{minor}  of $G(E,V)$ if  $H= G\setminus A / B$.  The graph $H(E',V')$  is a \textit{subgraph} of $G(E,V)$ if $V' \subseteq V$ and  $E' \subseteq E$.

 \begin{obs}\label{ContractPreservesCycles}
 The operation of contraction preserves cycles. That is, if $C$ is a cycle in $G$, then $C$ becomes a union of cycles in $G/A$.
 \end{obs}
 
 \begin{obs}\label{SubgraphIsMinor}
 Every subgraph (edge-induced,  vertex-induced or not) of $G$ is a minor of $G$. 
 \end{obs}

  Given a graph $G/A$, a  \textit{re-insertion}  of an edge   $e \in A$ consists of reversing  the contraction of  $e= \{v_{_{i}}, v_{_{j}}\} \in A$.    We write the re-insertion of  $e$ as $G/(A-e)$.   (We caution that  $A$ is a set, not a graph,  and $A - e$  is the set $A$ with the element $e$ removed from it).  Obviously, if $K$ is obtained from $G/A$  by  re-inserting some elements of $A$, then $K$ is a contraction-minor of $G$. And, if $G$ is an isthmus-less labelled  connected graph then $G/A$  is also an isthmus-less labelled connected graph. \\

\begin{example} \label{proofOfEquivalence}
\end{example}

 Let   $G(E,V)$ be a graph. A \textit{partial Hamiltonian cycle} of  $G(E,V)$  is a Hamiltonian cycle of a \textit{minor} of   $G(E,V)$ (See Appendix  \ref{exampleContraction}   for the definition of minors). Lemma  \ref{HCPSolution} claims  that if $C'$ is a partial hamiltonian cycle of  $G(E,V)$, then there is a contraction-minor $H$ of   $G(E,V)$ such that $C'$ is a hamiltonian cycle in $H$, and vice-versa. \\

 {\bf Proof of Lemma \ref{HCPSolution} }

{ \it Given a graph $G(E,V)$,  every  partial Hamiltonian cycle $G$  is a Hamiltonian cycle of a contraction-minor of $G$. 
}\\

{\bf Proof}

Suppose that $C'$ is  a hamiltonian cycle of the graph $H(E', V')$, where $H$ is a minor of $G(E,V)$.  Let $V= \{v_{_{1}}, v_{_{2}}, \cdots, v_{_{n}}\}$. Suppose that $V' = \{v_{_{1}}, v_{_{2}}, \cdots, v_{_{k}}\}$.  Let $C'= \{e_{_{1}}, e_{_{2}}, \cdots, e_{_{k}}\} $, where $k \leq n$.  Suppose that $C'$ is contained in a bigger partial Hamiltonian cycle $C=  \{e_{_{1}}, e_{_{2}}, \cdots, e_{_{k}},  e_{_{k+1}}, \cdots, e_{_{k+s}}  \} $, and  $C$ is Hamiltonian cycle of   a  contraction-minor of $G$, denoted $G/A$. Then contracting $G/A$ by   the edges $\{   e_{_{k+1}}, \cdots, e_{_{k+s}} \}$ yields the contraction-minor  $H$.    \\

Suppose that $C'$ is contained in a bigger partial Hamiltonian cycle $C=  \{e_{_{1}}, e_{_{2}}, \cdots, e_{_{k}},  e_{_{k+1}}, \cdots, e_{_{k+s}}  \} $  of a minor $H$, but $H$ is not   a  contraction-minor of $G$. That is   $H$ is obtained from $G$ by a series of deletions and contractions.  Let  $H= G\setminus A / B$.  We claim that $C$ is a Hamiltonian cycle of the graph $G/B$.  Indeed,   since the order of deletions and contractions is irrelevant,  suppose that $H$ is obtained from $G$ by contracting  all the edges in $B$, then deleting all the edges in $A$. If we now reverse the process by re-inserting the edges in $A$, we get the graph $G/ B$. But re-inserting all the edges in  $A$ does not create new vertices.  Hence  every edge re-inserted just creates a new cycle   connecting    the vertices already present. Thus, the cycle $C$ would still be present after re-inserting all the edges of $A$. Since $G/B$ and $G/B \setminus A$   have the same vertex-set we get that $C$ is a Hamiltonian cycle in $G/B$.\\

If $C'$ is not contained in any bigger   partial Hamiltonian cycle, then $C'$ is a cycle in $G(E,V)$.  Indeed, suppose that $C'$ is not a cycle in $G(E,V)$ but $C'$ is a cycle in $H(E', V')$, where $H$ is a minor of $G$.  Then,  $C'$ is  a non-cyclic  path in $G$ passing through the vertices of $V'$.  So, suppose  that $v_{_{1}}$ and $v_{_{k}}$ are not adjacent in $G$. But, since $v_{_{1}}$ and $v_{_{k}}$ are adjacent in $H$,    there must be a path $P$  from $v_{_{k}}$ to $v_{_{1}}$ in $G$  such that  contracting all the edges in $P$ yields the cycle $C'$.  Thus the $C' \cup P = C$  and  $C' \subset C$. This is a contradiction.     \\

Now, since $C'$ is a cycle in $G(E,V)$ and  $G(E,V)$ is isthmus-less, there are some other cycle $C''$, say, which contains some edges in $C'$ and some other edges not in $C'$.  For every such cycle $C''$, contracting all the edges not in $C'$  except one edge  yields a graph $H'(E', V')$, which is a contraction-minor of $G(E,V)$ and contains the Hamiltonian cycle $C'$.\\

\qed\\

{\bf Proof of Lemma  \ref{MisIndependentSolution}}\\
\textit{There is a bijection between the set of independent sets of vertices of  $G(E,V)$ and the set of  feasible sets of the  Maximal Independent Set Problem $\Pi(V, \gamma)$. That is, every  independent sets of vertices is a feasible set of the MISP, and  every feasible set of MISP is an independent sets of vertices. }\\

{\bf Proof}
Suppose that $Y$ is an independent set of vertices.  Consider the set   $ A=  \{v_{_{1}}, v_{_{2}}, \cdots, v_{_{k}}\}$, the set of all the vertices that are  not  connected to any vertex  in $Y$. Let $H$ be the graph obtained from $G$ by   contracting all  the edges connecting the vertices in $A$. That is, if there is a path $v_{_{i}}-e_{_{1}}-w_{_{1}}-e_{_{2}}- w_{_{2}}-\cdots-e_{_{k}}- w_{_{k}}-e_{_{k+1}}-v_{_{j}}$, where $v_{_{i}}, v_{_{j}} \in Y$ and $w_{_{1}}, \cdots,  w_{_{k}} \not \in Y$, then   contract all  the edges $e_{_{i}}$  for $2\leq i \leq k$.  Repeat the contractions recursively. We then get that  $Y$ is a solution of $\Pi(H,\gamma)$.  Thus,  every  independent set of  vertices of  the  graph $G(E,V)$ is a feasible set  of the  Maximal Independent Set Problem $\Pi(G[V], \gamma)$.\\

Conversely, let $Y$ be   feasible set of the MISP    $\Pi(G, \gamma)$. That is, there is a  contraction-minor $G/A$ such that  $Y$ is a solution of the problem $\Pi(G/A, \gamma)$.  Now, re-inserting the edges of $A$  preserves the fact that  no pair of  vertices in $Y$ are connected by an edge.  Hence $Y$ is an independent set of vertices  in the graph $G$. 
 
\qed\\


{\bf Proof of Lemma \ref{STPIndependentSolution}}

\textit{There is a bijection between the set of  independent sets  of edges of  $G(E,V)$ and the sets of  feasible sets of the  Spanning Tree Problem $\Pi(E, \gamma)$. That is, every tree  of $G$ is a feasible set of the Problem STP and  every  feasible set is a tree of $G$.
}\\

{\bf Proof}

Suppose that $Y$ is an independent set of edges. That is $A$ is a tree of $G$.   Consider the set of edges $ A= \{ e_{_{1}}, e_{_{2}}, \cdots, e_{_{k}}\}$  such that $Y \cup A$ is a spanning tree of $G$. Let $H$ be the graph obtained from $G$ by contracting the  edges  in $A$. Then $Y$ is a solution of $\Pi(H,\gamma)$.  Thus,  every  tree the  graph $G(E,V)$ is a feasible set  of the Spanning Tree Problem $\Pi(G[E], \gamma)$. \\

Conversely, let $Y$ be   feasible set of the STP  Problem   $\Pi(G, \gamma)$. That is, there is a  contraction-minor $G/A$, with  $ A= \{ e_{_{1}}, e_{_{2}}, \cdots, e_{_{k}}\}$ , such that  $Y$ is a solution of the problem $\Pi(G/A, \gamma)$.  Now, suppose that  re-inserting the edges of $A$  creates a cycle $C= \{ f_{_{1}}, f_{_{2}}, \cdots, f_{_{r}} \}  \subseteq Y$. Then, since the operation of contraction preserves cycles, we have that   $C_{_{1}} \subseteq Y$ in $G/A$, where $C_{_{1}}$ is a cycle from the union of cycles obtained by contraction from $C$, by Observation \ref{ContractPreservesCycles}. Thus $Y$ is not a solution of $\Pi(G/A, \gamma)$. A Contradiction.  \qed\\

Given a graph $G = (E,V)$, a \textit{matching} M is a set of edges with the property that no two of the edges have an end-vertex in common. A matching is \textit{maximum} if there is no matching of greater cardinality. In particular, a maximum matching is called \textit{perfect} if every vertex of G is matched. The \textit{Matching Problem} consist of finding a maximum matching of the graph $G = (V,E)$.  \\

{\bf Proof of Lemma \ref{MaximumMatchingSolution}}

\textit{There is a bijection between the \textit{matchings}  of   $G(E,V)$ and the sets of   feasible sets of the  Maximum Matching   Problem $\Pi(G[E], \gamma)$. That is, every matching  of $G$ is a feasible set of the Problem MMP and  every  feasible set is a matching of $G$.  
}\\

{\bf Proof}

Let $ Y$ be a matching of the graph $G$. That is, $Y= \{ e_{_{1}}, e_{_{2}}, \cdots, e_{_{s}}\}$  is a set of edges of $G$ such that  no pair of edges in $Y$   shares the same end-vertex $v$. Let $v_{_{i,1}}$ and $v_{_{i,2}}$ be the end-vertices of the edges $e_{_{i}}$.  Let $V_{_{Y}}$ be  the set of all vertices $v_{_{i,1}}, v_{_{i,2}}$  for all edges in $Y$. Now,  if there is a path $v_{_{i}}-e_{_{1}}-w_{_{1}}-e_{_{2}}- w_{_{2}}-\cdots-e_{_{k}}- w_{_{k}}-e_{_{k+1}}-v_{_{j}}$, where $v_{_{i}}, v_{_{j}} \in V_{_{Y}}$ and $w_{_{1}}, \cdots,  w_{_{k}} \not \in V_{_{Y}}$, then   contract all  the edges $e_{_{i}}$  for $2\leq i \leq k+1$, and repeat the process recursively  to obtain  the graph $G/A$.  Thus,  the graph $G/A$ contains only vertices that are the end-points of edges in $Y$.  Thus $Y$ is a maximum matching in $G/A$  since $Y$ is a \textit{Perfect Matching}  in $G/A$  as it covers all its  vertices. \\
  
  Conversely, let $Y$ be  a  feasible set of the MMP  Problem   $\Pi(G, \gamma)$. That is, there is a  contraction-minor $G/A$ such that  $Y$ is a maximum matching  of the graph $G/A$.  Now, re-inserting the edges of $A$  preserves the fact that  no pair of  edges in $Y$ share the same end-vertex.  Hence $Y$ is a matching in the graph $G$. 
         
\qed\\

\begin{example} \label{exampleSTP1} 
\end{example}


\begin{figure}[h]
\center
\includegraphics[scale=0.3]{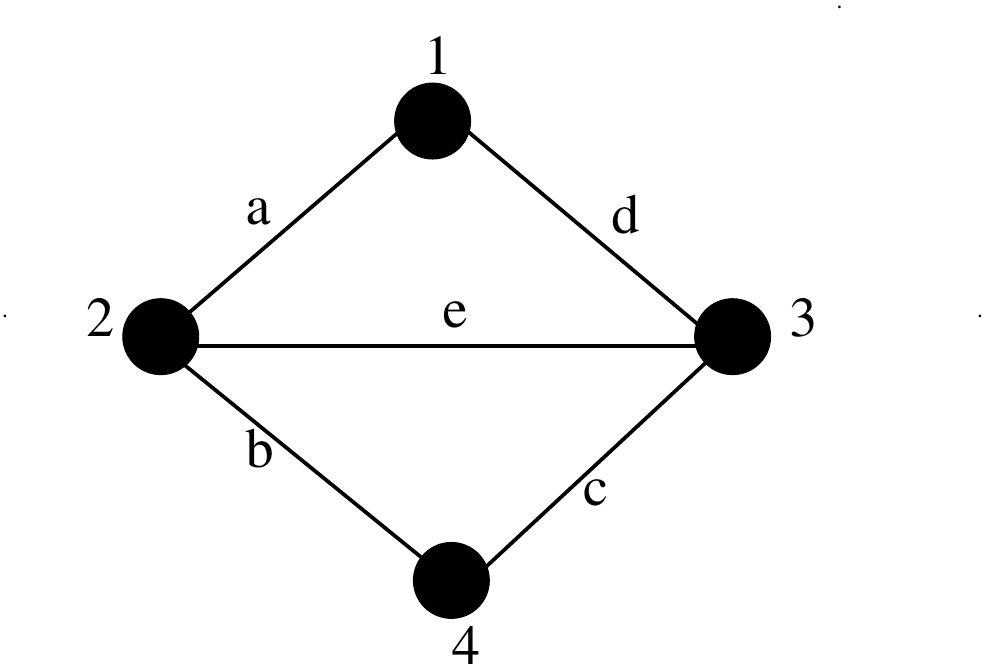}
\caption{A graph $G$.}
\label{favoriteGraph}
\end{figure}   
 

 Consider the graph $G$ given in Figure \ref{favoriteGraph}.   Let ${\Pi}$  be the problem, denoted STP, that consists of finding a spanning tree of $G$. That is, finding a set $Y$ of edges that connects all the vertices of $G$, but   does not contain  a cycle. Thus $X= E$, the set of edges of $G$.  Let $\mathcal{I}$  denote the set of all feasible sets of $\Pi$.  We have 
 
 \[
 \mathcal{I}= \{ \textrm{all the spanning trees of $G$,  all the sub-trees of $G$, the empty set.} \}
 \]

Indeed, we have that solutions or bases  of $\Pi$ are spanning trees of $G$. And sub-solutions of $\Pi$ are the sub-trees of $G$. For, let $Y'$  be a subtree of $G$. Then there is a set of edges $A$ such that $Y' \cup A$  is a spanning tree of $G$. Thus, $G/A$ is a contraction-minor of $G$ such that $Y'$ is a solution of $\Pi$  restricted to $G/A$.  For example, $\emptyset$ is a sub-solution since $\emptyset$ is the solution of $\Pi$ restricted to $G / \{a, b, c\} $. The singletons $\{a\}$ and  $\{b\}$ are sub-solutions since they are  solutions of $\Pi$ restricted to $G/ \{c,d\}$. The singletons $\{c\}$, $\{d\}$ and $\{e\}$ are sub-solutions since they are  solutions (spanning trees) of $\Pi$ restricted to $G/ \{a,b\}$. All the two-set subsets are also sub-solutions. For example,  the sets $\{a,e\}$, $\{a,d\}$,  $\{a,b\}$, $\{d,b\}$,$\{d,e\}$ are  sub-solutions since they are  solutions (spanning trees) of $\Pi$ restricted to $G/ \{c\}$. 
All the three-set subsets except for   $\{a,d,e\}$  and $\{b,c,e\}$ are bases. That is, they are solutions of  $\Pi(G)$. Thus,   a feasible set  is any subset  of $\{a,b, c,d,e \}$ that does not contain a cycle.  Hence, for the STP problem,  $\mathcal{I}$  is the set of feasible sets of the cycle matroid  of $G$. \\

As an acid-test, this example shows how our definition of feasible sets is a natural  extension of the definition of feasible sets of greedoids.  More generally, the present paper  aims at showing   that, if a problem can be solved in polynomial time, or a solution can be checked in polynomial time, then  a solution of every  such a search problem  is a `basis'  of some `greedoid-like' combinatorial  structure.  \\ 
 
Notice that, in the STP example, a subset may be a solution  for many sub-instances.  For example, $\{a,b\}$  is a solution for $G/c$, $G/d$ or $G/e$. And a sub-instance may have many possible solutions. However, in Definition 2.2, we only require the existence of  one sub-instance $X'$ such that $\Pi(X')=Y'$ for $Y'$ to be a feasible set of $\Pi$. \\

  The sub-solutions of $\Pi$ are the solutions of $\Pi$ instanced on minors   $G/ B$, where $B \subset E$. For example, $\emptyset$ is a sub-solution since $\emptyset$ is the solution of $\Pi$ restricted to $G / \{a, b, c\} $. And,  $cl(\emptyset)= G/ \{a, b, c\}$. The set $\{a,b\}$ is also a sub-solution since $\{a,b\}$ is the solution of $\Pi$ restricted to $G/ \{c\}$. And,  $cl(\{a,b\})= G / \{c\}$.\\  
  
   Notice that, for this example, our notion of closure corresponds with the notion of closure in Matroids, defined as follows. For all subsets $X' \subseteq X$, let the rank of $X'$ be a function  $r: 2^X \rightarrow \mathbb{N}^+$ (positive integers),   defined as

\[
r(X')= |Y'|,
\]

where $Y'$ is the largest element of $\mathcal{I}$ contained in $X'$. For all subsets $X'$, let the closure of $X'$, denoted $cl(X')$, be defined as 

\[
cl(X')= \{ e \in X: r(X' \cup e)= r(X')\}.
\] 
 
    Now, each sub-solution $Y'$ is a sub-tree of the graph $G$, and  $cl(Y')$ is just the contraction-minor spanned by $Y'$. That is, $cl(Y')$ is the set of edges that do not increase the rank of the sub-tree $Y'$.\\
    
     It is part of the folklore of Matroid Theory that the family of all the feasible sets of the STP is the family of all the spanning trees and sub-trees of the graph $G$. And this is the family of the feasible sets of a  matroid, the cycle matroid of $G$. Hence, the set system $(X, \mathcal{I})$ of STP defines  a matroid..

\qed\\

\begin{example} \label{exampleMIS}
\end{example}

\begin{figure}[h]
\center
\includegraphics[scale=0.3]{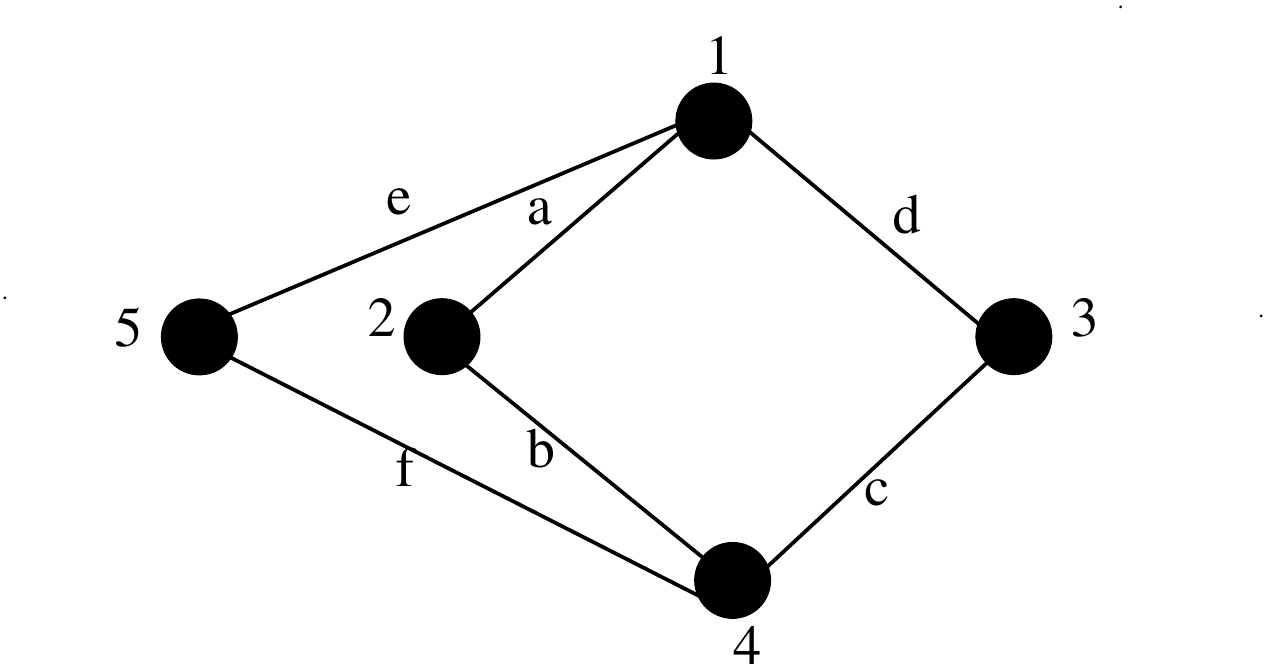}
\caption{A graph $G$.}
\label{Mis}
\end{figure} 

  
Consider the graph  $G$ given in Figure \ref{Mis}. Let ${\Pi(G)}$, denoted MIS,  be the problem consisting of finding a maximal independent set of vertices  of $G$. That is, finding a set $Y$ of vertices that are not adjacent to each other and no other vertex can be added without violating independence.   Thus $X= V$, the set of vertices of $G$.  A solution (basis) would be the set

\[
Y= \{1,4 \}. 
\]










We have 

\[
\mathcal{I}= \{ \emptyset, \{1\},\{2\},\{3\},\{4\}, \{5\}, \{1,4\}, \{2,3\}, \{2,5\}, \{3,5\}  , \{2,3,5\}  \}
\]

Indeed,  $\emptyset$ is a solution of $\Pi$  restricted to   $G/\{a,b,c,d,e,f\}$, while  $\{1\}$ or  $\{4\}$  are solutions of $\Pi$ restricted to $G/\{a\}$, and  $\{2,5\}$ is a solution of  $\Pi$ restricted to $G/\{c\}$.\\

Another maximal independent set (basis) of $\Pi(G)$ is the set $\{2,3,5\}$.  It is worth noticing that, although, by Definition 2.2,      $\{1,4\}$ and $\{2,3,5\}$ are both bases, they have not the same cardinality. Hence, the set system $(X, \mathcal{I})$  associated with  MIS can not be that of a greedoid. However,  one may check that every  feasible set  that is not a basis is augmentable. For example, $\emptyset$ can be augmented to become any singleton. The sets $\{1\}$ or $\{4\}$ can be augmented into $\{1,4\}$. This is an illustration of the  result given in Lemma \ref{MISAugmentable}, which, along with Lemma \ref{MISAccessible},  is much used in the present paper. \\

Since the closure of  the Maximal Independent Set Problem (MIS) is much used in the  proof of the main theorem, we give a full description of it. For the MIS  Problem,   $cl(Y')= Y' \cup  A$, where $A$  is the set of all  the vertices in $X -Y'$ that are connected to some vertex of $Y'$ in $G$. Hence, $cl(Y')$ is unique (as a set of vertices)  for all feasible sets $Y'$.\\

  As a contraction-minor, $G[cl(Y')]$ can be constructed from  the graph $G$ by merging all the vertices not adjacent to $Y'$  with vertices adjacent to $Y'$   as follows:\\

\begin{figure}[h]
\center
\includegraphics[scale=0.42]{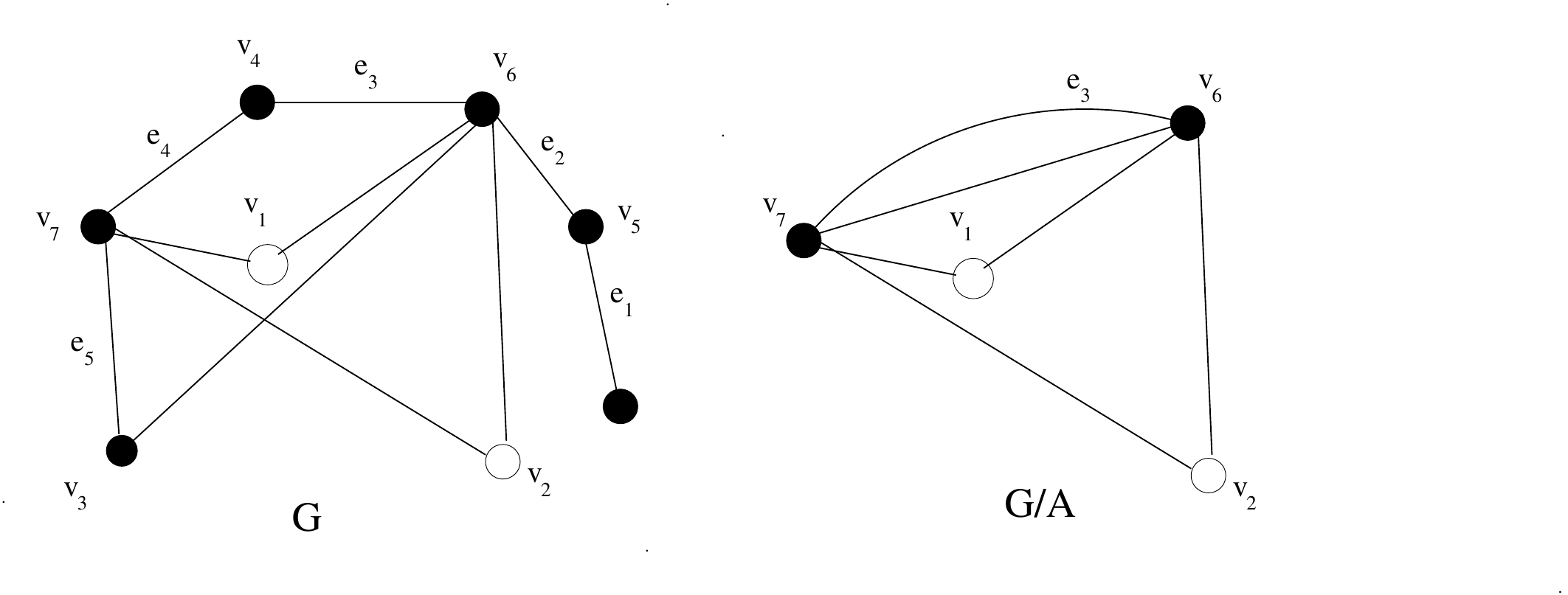}
\caption{A closure of the feasible set  $\{v_{_{1}},  v_{_{2}}\} $, denoted $G/A$,  obtained from the graph $G$   by the contraction of the edges $e_{_{1}}$, $e_{_{2}}$, $e_{_{4}}$ and  $e_{_{5}}$. Notice that all these contractions merge vertices that are not adjacent to $v_{_{1}}$ or $v_{_{2}}$ with vertices that are  adjacent to $v_{_{1}}$ or $v_{_{2}}$.   }
\label{MISClosure}
\end{figure}

- If  the set  $\{(v_{_{0}}, v_{_{1}}),  (v_{_{1}}, v_{_{2}})    \cdots,(v_{_{k-1}}, v_{_{k}})\}$ is a path of  edges of $G$  such that vertices  $v_{_{0}}, v_{_{1}}, \cdots, v_{_{k}}$  are  not in $Y'$,    the vertex $v_{_{0}}$ and $v_{_{k}}$    are connected to vertices in $Y'$, the vertices   $v_{_{1}}, v_{_{2}}, \cdots, v_{_{k-1}}$  are not connected to vertices in $Y'$,     then  merge the vertices $v_{_{0}}$  and $v_{_{k-1}}$ by contracting all the   edges $(v_{_{0}}, v_{_{1}}),  (v_{_{1}}, v_{_{2}})    \cdots,(v_{_{k-2}},v_{_{k-1}})$.  Repeat the process recursively. Thus,   every such path is reduced to the edge   $(v_{_{0}}, v_{_{k}})$,  with possibly some loops, as illustrated in Figure \ref{MISClosure} (contraction of the edge $e_{_{4}}$ ).  \\

- If  the set  $\{(v_{_{0}}, v_{_{1}}),  (v_{_{1}}, v_{_{2}})    \cdots,(v_{_{k-1}}, v_{_{k}})\}$ is a path of  edges of $G$  such that vertices  $v_{_{0}}, v_{_{1}}, \cdots, v_{_{k}}$  are  not in $Y'$,    the vertex $v_{_{0}}$    is connected to vertices in $Y'$, the vertices   $v_{_{1}}, v_{_{2}}, \cdots, v_{_{k}}$  are not connected to vertices in $Y'$,     then obtain $G[cl(Y')]$  from $G$  by merging the vertices $v_{_{0}}$  and $v_{_{k}}$ by contracting all the   edges $[(v_{_{0}}, v_{_{1}}),  (v_{_{1}}, v_{_{2}})    \cdots,(v_{_{k-1}}, v_{_{k}})]$.  Repeat the process recursively.  Thus,  every such path is reduced to a single vertex  $v_{_{0}}$  with possibly some loops, as illustrated in Figure \ref{MISClosure} (contraction of the edge $e_{_{1}}$  or $e_{_{2}}$ ). \\

 One may check that this definition also makes it unique as a contraction-minor. It is   also maximal since any one contraction less leaves a vertex $v_{_{i}}$  that is not connected to $Y'$. \\

\begin{example} \label{exampeMisMax}
\end{example}
\begin{figure}[h]
\center
\includegraphics[scale=0.3]{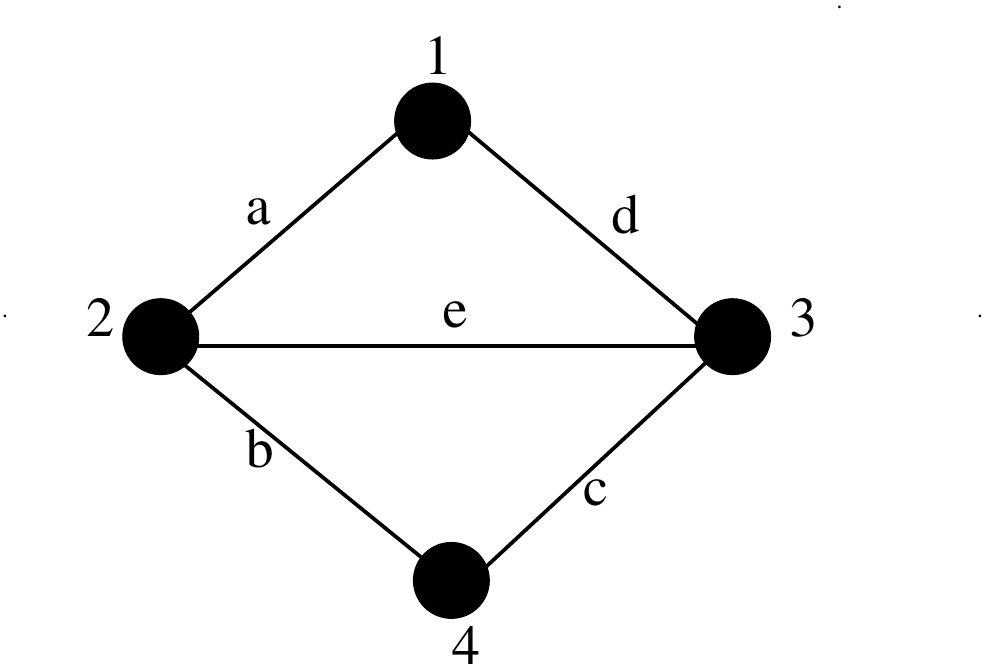}
\caption{A graph $G$.}
\label{MisMax}
\end{figure}   

Consider a search  problem $\Pi$ on  the graph given in Figure \ref{MisMax}. Notice that  the bases (solutions) of  the MIS problem on the graph of Figure  \ref{MisMax}  are the sets of vertices $B_{_{1}}, B_{_{2}}, B_{_{3}}$, where  

\[
B_{_{1}}= \{1,4 \},  B_{_{2}}= \{2\}, B_{_{3}}= \{3\} 
\]

However,  the \textit{Maximum Independent Set } problem, denoted $MaxIS$, consists of finding an independent set of the greatest cardinality. The bases (solutions) for this problem is $B_{_{1}}$  only. The sets $B_{_{2}}= \{2\}$ and  $B_{_{3}}= \{3\}$ are not bases for the $MaxIS$, but there are sub-solutions. Indeed, they are solutions when the instance is the contraction-minor $G/c$. It is worth noticing that in $MaxIs$, the sub-solutions $B_{_{2}}$  and $B_{_{3}}$  are not augmentable.

\vspace{3mm}

\begin{example} \label{exampleMATCHING} 
\end{example}

 Let $\Pi(G[X], \gamma) $, where $X$ is the set of edges of $G$, consist of finding a  maximum matching of the graph in Figure \ref{MaxMatch}.  The  matchings  are given as follows:

\begin{figure}[h]
\center
\includegraphics[scale=0.5]{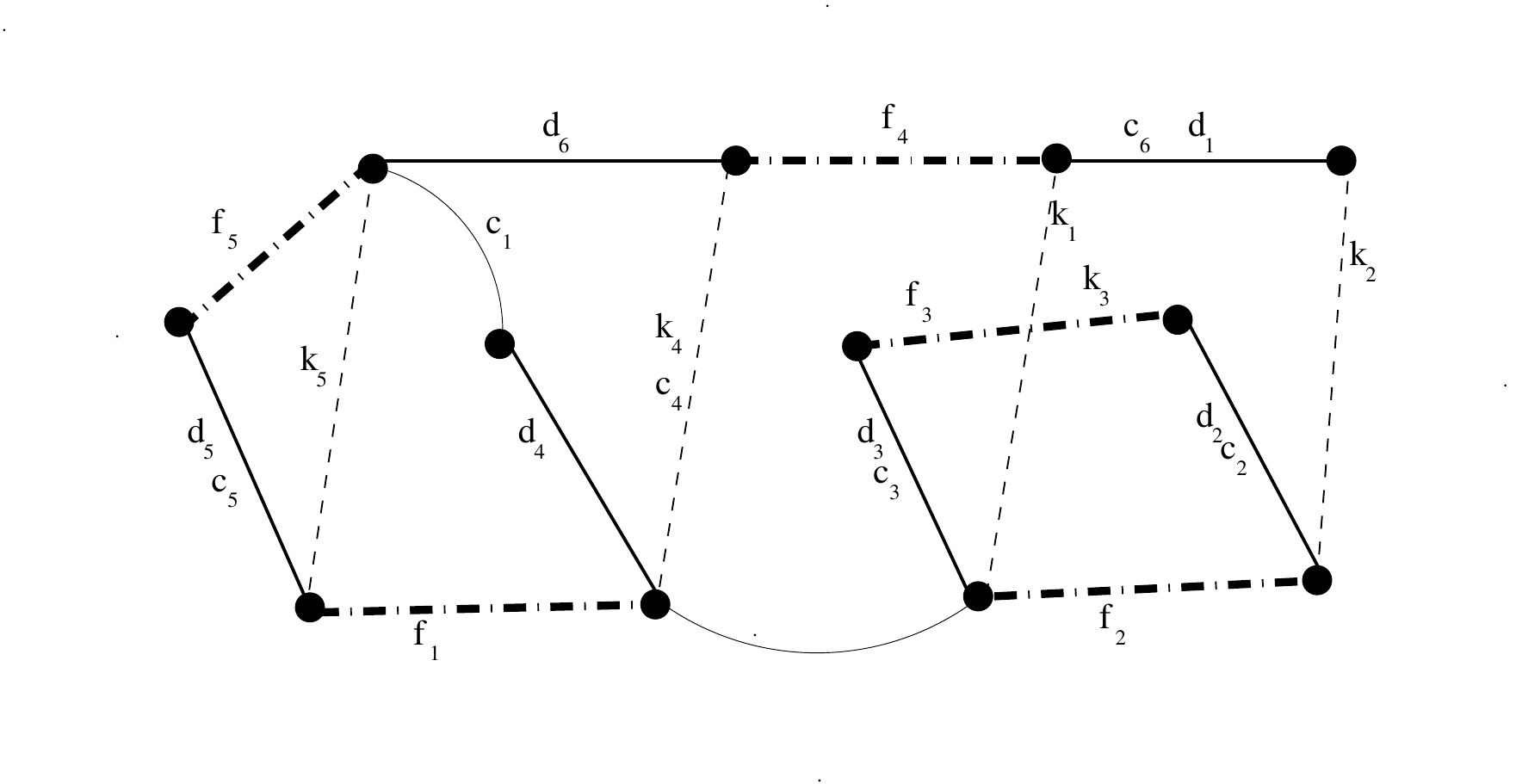}
\caption{Graph G}
\label{MaxMatch}
\end{figure}

$C= \{ c_{_{1}}, c_{_{2}}, c_{_{3}}, c_{_{4}}, c_{_{5}}, c_{_{6}} \}$\\
$D= \{ d_{_{1}}, d_{_{2}}, d_{_{3}}, d_{_{4}}, d_{_{5}}, d_{_{6}} \}$\\
$F= \{ f_{_{1}}, f_{_{2}}, f_{_{3}}, f_{_{4}}, f_{_{5}}\}$\\
$K= \{ k_{_{1}}, k_{_{2}}, k_{_{3}}, k_{_{4}}, k_{_{5}} \}$\\

The matchings $C$ and $D$ are maximum, while  the matchings $F$ and $K$ are not maximum. So, the solutions of the problem $\Pi(G[X],\gamma)$ are the matchings $C$ and $D$. Moreover the matching $F$ and $K$ are not fast-augmentable. That is, there  is no edge $w$ such that $F \cup w$ or $K \cup w$ is a matching. \\

  Let  $Y'$ be a feasible set of the Maximum Matching Problem  $\Pi(G[X], \gamma) $. By Lemma \ref{MaximumMatchingSolution}, $Y'$ is a matching in $G[X]$.  Let $G/A$ be the contraction-minor of $G$ such that $Y'$ is a maximum matching in $G/A$, with $A \geq \emptyset$.    A closure of $Y'$  is $Y' \cup E$, where $E$ is the subset of  all the edges $e$  of  $A$  such that $e$ shares the same vertex with some edge in $Y'$.  That is, re-inserting the edge  $e \in E$ does not  create a matching  of bigger cardinality $Y' \cup e$.

\begin{lemma}
The problem MMP is accessible. 
\end{lemma}

{\bf Proof:}

Let $Y= \{ e_{_{1}}, e_{_{2}}, \cdots , e_{_{n}} \} $ be a maximum matching in the graph $H= G/A$, with $A \geq \emptyset $, and consider the edge $e_{_{1}}$.  Suppose that $Z= \{ f_{_{1}}, f_{_{2}}, \cdots , f_{_{n}} \}$  is another   maximum matching in the graph $H= G/A$.  Then   $e_{_{1}}$ must share some end-vertex with some  edges in $Z$, lest $Z$ is not a maximum matching in $H$ since $e_{_{1}}$ can be added to it.  Hence,  three cases may arise:

(1) $e_{_{1}}= f_{_{j}}$ for some  $j$\\
(2) There is a path $(f_{_{j}},e_{_{1}},f_{_{k}})$\\
(3) There is a path $(f_{_{j}},e_{_{1}}, g)$, where the edge $g \not \in Z$, and $e_{_{1}}$  does not share a  vertex with another edge $f_{_{k}}$, with $k\not = j$.  \\

 Consider the graph  $H/B$, where $B$ contains the edge $e_{_{1}}$ and all the edges $g$ if  there is a path $(f_{_{j}},e_{_{1}}, g)$, where the edge $g \not \in Z$.  Then $Y$ is a maximum matching in the graph $H/B$. Indeed,  if case (1)  arises, then $Z-e_{_{1}}$ has the same cardinality as $Y-e_{_{1}}$. If  case (2) arises,  then $Z$ looses one element as $Y$ looses the element $e_{_{1}}$. If case (3) arises, $Z$  is a maximum matching in $H/e_{_{1}}$, while $Y- e_{_{1}}$ is not a maximum matching in $H/e_{_{1}}$. However, contracting by $g$ makes $Z$ to loose one element. Hence $Y- e_{_{1}}$ is  a maximum matching in $H/e_{_{1}}/g$. 

\qed
  
\begin{example} \label{exampleHCP} 
\end{example}

 Let $\Pi$ consist of finding a  Hamiltonian cycle of the graph in Figure \ref{MisMax}. The set of edges $ C=\{a,b,c,d \}$ is a  solution  (basis), since it is a  Hamiltonian cycle of the graph $G$. The set of edges $ C_{_{1}}=\{a,b,d\}$ is a sub-solution of $\Pi$, since it  is a Hamiltonian cycle for the sub-instance $G/ \{c\}$.  Notice also that the set of edges $C_{_{2}}=\{a,d,e\}$ is another Hamiltonian cycle of the graph $G/ \{c\}$. We have,

\begin{eqnarray*}
\mathcal{I} &=& \{\emptyset, \textrm{all the singletons},\textrm{all the 2-subsets}, \{a,d,e\}, \{b,c,e\},\\
 && \{a,b,c\}, \{a,b,d\}, \{b,c,d\},\{a,c,d\}, \{a,b,c,d\}\}
\end{eqnarray*}
 
 Indeed, the feasible set $ B= \{a,b,c,d\}$ is the unique basis. Moreover, consider any subset   $C \subseteq B$. Then $B - C$ is a Hamiltonian cycle of the graph $G / C$. Hence,  all the subsets of $B$ are feasible sets. 
 Now, consider the feasible sets that are not subsets of $B$. The singleton  $\{e\}$ is a   Hamiltonian cycle of the graph $G / \{b,c,d\}$. The 2-subsets $\{ a,e\}$, $\{ c,e\}$, $\{ b,e\}$ and $\{d,e\}$  are Hamiltonian cycles of the graph $G/ \{b,d\}$, $G/ \{b,d\}$, $G/ \{a,c\}$ and $G/ \{a,c\}$, respectively. The 3-subsets $\{a,d,e\}$ and $\{a,d,c\}$ are Hamiltonian cycles of the graphs $G/ \{b\}$, while $\{b,c,e\}$ and $\{b,c,d\}$ are Hamiltonian cycles of the graph $G/ \{a\}$. One may check that the set system $(X, \mathcal{I})$ of the Hamiltonian Cycle Problem does not define a greedoid. \\
 
 For the Hamiltonian Cycle Problem and a feasible set $Y'$,  the graph  $G[cl(Y')]$ is obtained from $G$ as follows: \\

\begin{figure}[h]
\center
\includegraphics[scale=0.42]{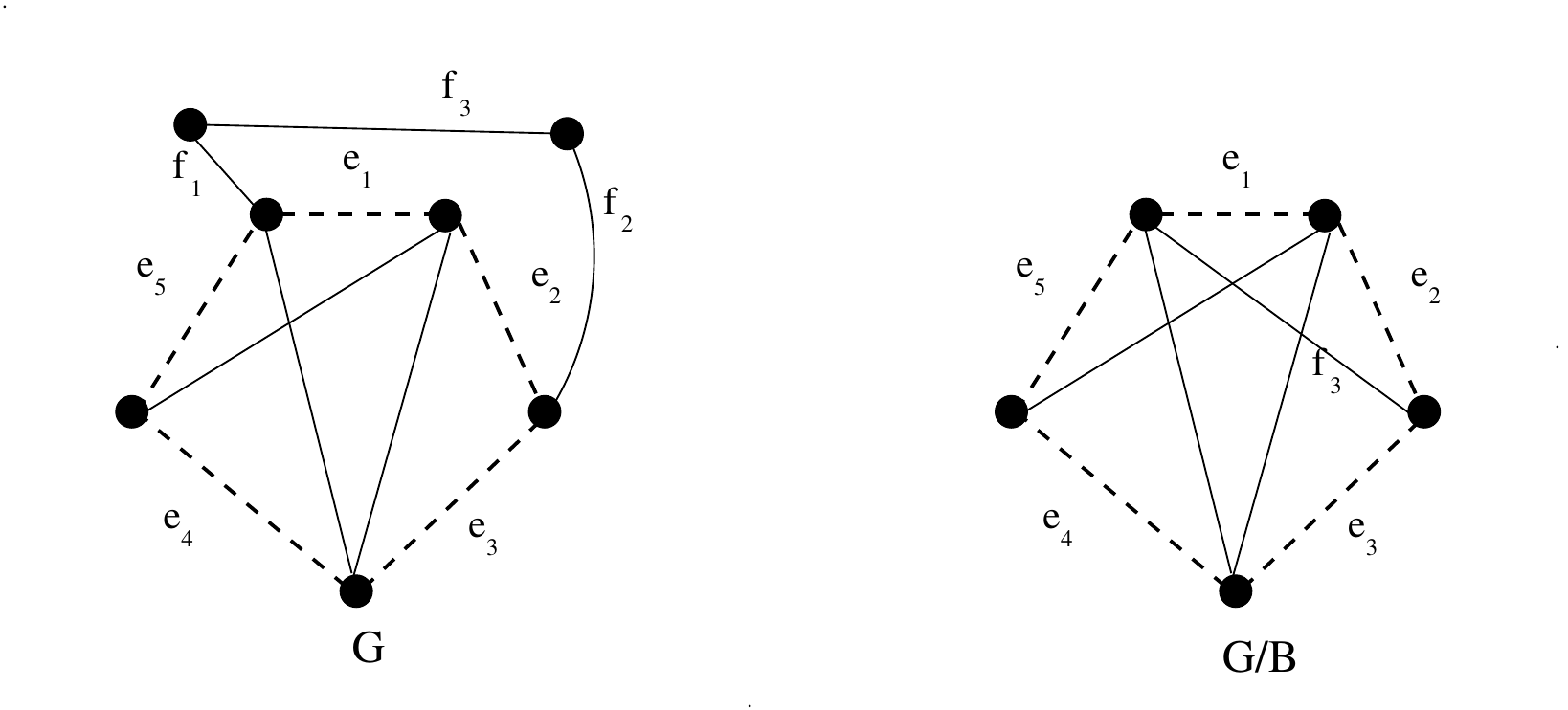}
\caption{A closure of the feasible set  $\{e_{_{1}},  e_{_{2}},  e_{_{3}}, e_{_{4}}, e_{_{5}}\} $, denoted $G/B$,  obtained from the graph $G$   by the contraction of the edge $f_{_{1}}$ and $f_{_{2}}$. }
\label{HamiltonianClosure1}
\end{figure}

-  Let $ Y'= \{ e_{_{1}}, e_{_{2}}, \cdots e_{_{k}}\}$ be a feasible set. That is,  $Y'$ is a Hamiltonian cycle of a contraction-minor $G/A$.  If $Y'$ is a cycle in $G$, then one obtains $G[cl(Y')]$ by contracting a minimal number of edges $e$ such that all the vertices $v$ not incident to  edges in $Y'$ are merged with vertices $w$ incident to edges in $Y'$, as illustrated in Figure \ref{HamiltonianClosure1}. \\

If $Y'$ is  not a cycle in $G$, then one obtains $G[cl(Y')]$ by contracting a minimal number of edges $e$  to turn $Y'$ into a cycle, then one contracts a minimal number of edges   such that all the vertices $v$ not incident  to edges  in $Y'$ are merged with vertices $w$ incident to edges in $Y'$.

\begin{figure}[h]
\center
\includegraphics[scale=0.42]{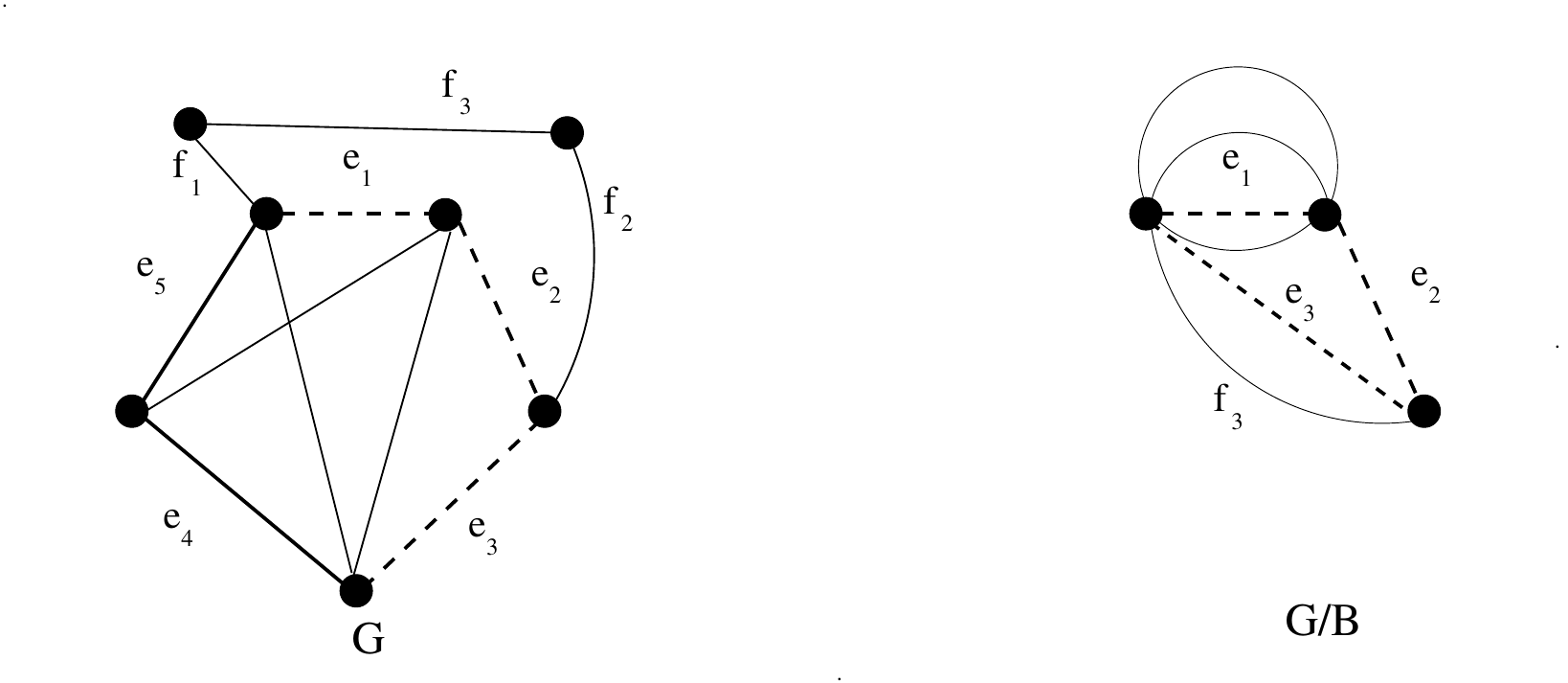}
\caption{A closure of the feasible set  $\{e_{_{1}},  e_{_{2}},  e_{_{3}}\} $, denoted $G/B$,  obtained from the graph $G$   by first  contracting  the edges $e_{_{5}}, e_{_{4}}$  to turn  $\{e_{_{1}},  e_{_{2}},  e_{_{3}}\} $ into a cycle, then contracting  $f_{_{1}}$ and $f_{_{2}}$ to turn $\{e_{_{1}},  e_{_{2}},  e_{_{3}}\} $ into a Hamiltonian cycle.  }
\label{HamiltonianClosure2}
\end{figure}

 \vspace{3mm}

 
 
   

\vspace{3mm}


\pagebreak



{\bf Acknowledgments}
The author  is grateful to the various universities where he spent his academic life. He is especially grateful to University of Oxford where he was introduced to Matroid Theory by Prof.  Dominic Welsh. He is very grateful to the University of Qatar where he did most of the work for this paper. He is finally very grateful to the University of London, Queen Mary College, where he was introduced to Graph Theory by Prof. David Arrowsmith and Prof. Peter Cameron. \\

\pagebreak

\end{document}